\documentclass[aps,pra,twocolumn,showpacs,amsmath,amssymb,superscriptaddress,reprint]{revtex4-1}

\usepackage[utf8]{inputenc}

\usepackage{graphicx}
\usepackage[colorlinks=true,citecolor=blue]{hyperref}
\usepackage{units}

\usepackage{verbatim} 

\newcommand{\ket}[1]{|#1\rangle}
\renewcommand{\vec}[1]{\mathbf{#1}}
\newcommand{\kBT}{k_\text{B}T}

\def\up{\uparrow}
\def\down{\downarrow}

\begin{document}

\title{Josephson response of a conventional and a noncentrosymmetric superconductor coupled via a double quantum dot}

\author{Bj\"orn Sothmann}
\affiliation{D\'epartement de Physique Th\'eorique, Universit\'e de Gen\`eve, CH-1211 Gen\`eve 4, Switzerland}
\author{Rakesh P. Tiwari}
\affiliation{Department of Physics, University of Basel, Klingelbergstrasse 82, CH-4056 Basel, Switzerland}

\date{\today}

\begin{abstract}
We consider transport through a Josephson junction consisting of a conventional s-wave superconductor coupled via a double quantum dot to a noncentrosymmetric superconductor with both, singlet and triplet pairing. We calculate the Andreev bound state energies and the associated Josephson current. 
We demonstrate that the current-phase relation is a sensitive probe of the singlet-triplet ratio in the noncentrosymmetric superconductor. 
In particular, in the presence of an inhomogeneous magnetic field the system exhibits a $\varphi$-junction behavior.
\end{abstract}


\maketitle

\section{Introduction}
Conventional s-wave superconductivity is well understood by Bardeen-Cooper-Schrieffer (BCS) theory in terms of the formation of spin-singlet Cooper pairs~\cite{bardeen_theory_1957}. Recently, there has been a growing interest in a new class of superconducting materials termed noncentrosymmetric superconductors (NCSCs) whose unit cell lacks inversion symmetry~\cite{bauer_heavy_2004,kimura_pressure-induced_2005,sugitani_pressure-induced_2006,honda_pressure-induced_2010,akazawa_pressure-induced_2004,
togano_superconductivity_2004,badica_superconductivity_2005,yuan_$s$-wave_2006,bauer_unconventional_2010,karki_structure_2010,amano_superconductivity_2004,
butch_superconductivity_2011,joshi_superconductivity_2011,tafti_superconductivity_2013}. Consequently, these materials exhibit a strong antisymmetric spin-orbit interaction. In the superconducting state, the spin-orbit interaction gives rise to a mixing of spin-singlet and spin-triplet pairing~\cite{gorkov_superconducting_2001,bauer_non-centrosymmetric_2012}.

An important open question in the field of NCSCs is to determine the precise value of the singlet-triplet ratio in a given material. 
By now, there are different proposals based on steps in the current-voltage characteristics of a NCSC-NCSC junction~\cite{borkje_tunneling_2006,borkje_using_2007}, the formation of Andreev bound states and associated zero-bias anomalies~\cite{iniotakis_andreev_2007,vorontsov_surface_2008}, and the peak structure of the nonlocal conductance due to crossed Andreev reflections between a NCSC and spin-polarized normal metals~\cite{fujimoto_unambiguous_2009}.
Furthermore, the low-temperature behavior of the critical current~\cite{asano_josephson_2011} and the occurrence of higher harmonics in the current-phase relation~\cite{rahnavard_magnetic_2014} in NCSC-NCSC Josephson junctions have been suggested as probes of the singlet-triplet ratio. For a Josephson junction between a conventional superconductor and a NCSC a transition between 0 and $\pi/2$ junction behavior has been predicted as a function of the singlet-triplet ratio~\cite{klam_josephson_2014}. Furthermore, the occurrence of a two-peak structure in Raman spectra has been discussed as a signature of the singlet-triplet ratio~\cite{klam_electronic_2009}.

\begin{figure}
	\includegraphics[width=\columnwidth]{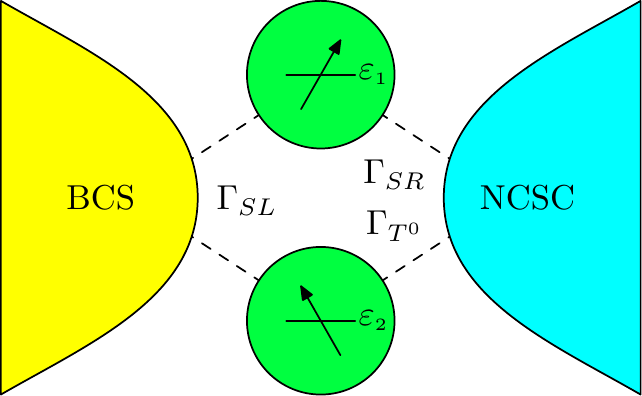}
	\caption{\label{fig:model}(Color online) Schematic model of our setup. A double quantum dot (green) subject to an inhomogeneous magnetic field is tunnel coupled to a conventional $s$-wave  superconductor (yellow) on the left as well as to a noncentrosymmetric superconductor (light blue) on the right.}
\end{figure}
Here, we propose a Josephson junction consisting of a conventional $s$-wave superconductor and a NCSC coupled via a double quantum dot subject to an inhomogeneous magnetic field (see Fig.~\ref{fig:model}) as a sensitive probe of the singlet-triplet ratio in the NCSC. A similar Josephson junction consisting of two singlet $s$-wave superconductors in the absence of an inhomogeneous magnetic field gives rise to a spin-dependent Josephson current~\cite{choi_spin-dependent_2000}. Double quantum dots tunnel-coupled to an unconventional superconductor like Sr$_2$RuO$_4$ can provide signatures of chiral edge states associated with it~\cite{tiwari_quantum_2014}. Recently, double quantum dots coupled to an $s$-wave superconductor have been shown to convert spin-singlet correlations into spin-triplet correlations and vice versa in the presence of an inhomogeneous magnetic field in a controlled and tunable way~\cite{sothmann_unconventional_2014}. It is this conversion between singlet and triplet Cooper pairs that allows the probing of both 
singlet and triplet pairings in the NCSC by detecting the Josephson current in the proposed setup. Thus, Josephson junctions based upon double quantum dots can be very useful for investigation of the symmetry of the superconducting order parameters. Compared to a tunnel barrier without quantum dots~\cite{klam_josephson_2014}, our setup offers the advantage of increased tunability as level positions in a quantum dot can easily be shifted by a gate voltage. Furthermore, our setup makes use not only of a homogeneous magnetic field but also of an inhomogeneous field component.

In particular, we find that the interplay between the singlet and triplet components can give rise to a $\varphi$-junction behavior, i.e., a finite phase difference $\varphi$ in the ground-state as well as a finite Josephson current at zero phase difference. Such junctions have been predicted to occur in ferromagnetic Josephson junctions with spin-orbit interactions~\cite{buzdin_direct_2008}, in Josephson junctions containing a multilevel quantum dot with spin-orbit coupling and a magnetic field~\cite{brunetti_anomalous_2013}, in Josephson junctions with spin-orbit nanowires subject to magnetic fields~\cite{yokoyama_anomalous_2014}, in multiterminal quantum-dot Josephson junctions subject to frustration between $0$ and $\pi$ states~\cite{feinberg_spontaneous_2014}, and in complex ferromagnetic Josephson junctions~\cite{buzdin_periodic_2003,pugach_method_2010,goldobin_josephson_2011,
kulagina_spin_2014}. In the latter type of setup, the occurrence of a $\varphi$ junction has recently been confirmed experimentally~\cite{sickinger_experimental_2012}. In this work, we suggest the occurrence of a $\varphi$ junction as a sensitive tool to probe the singlet-triplet ratio of the NCSC. We demonstrate that apart from extracting this information directly from the current-phase relationship we can also obtain information about the ratio from the difference between backward critical current and forward critical current.

The paper is organized as follows. In Sec.~\ref{sec:model} we introduce our theoretical model of the Josephson junction. We discuss the Andreev bound state energies and the associated Josephson current in Sec.~\ref{sec:results} and give conclusions in Sec.~\ref{sec:conclusions}.

\section{\label{sec:model}Model}
We consider a double quantum dot embedded in a Josephson junction formed by a conventional $s$-wave superconductor as well as a NCSC (cf. Fig.~\ref{fig:model}). The total Hamiltonian describing our junction is given by
\begin{equation}
 H=H_\text{BCS}+H_\text{NCSC}+H_\text{DQD}+K_\text{BCS}^\text{DQD}+K_\text{NCSC}^\text{DQD}.
\end{equation}
Here,
\begin{equation}
 H_\text{BCS}=\sum_{\vec k,\sigma}\epsilon^\text{BCS}_{\vec k}d^{\dagger}_{\vec k\sigma}d^{}_{\vec k\sigma}-\Delta_\text{BCS} \sum_{\vec k}(e^{i\Phi}d^{}_{-\vec k\downarrow}d^{}_{\vec k\uparrow} +\text{H.c.}),
\end{equation}
describes a conventional BCS superconductor with dispersion $\epsilon^\text{BCS}_{\vec k}$, superconducting energy gap $\Delta_\text{BCS}$, and phase $\Phi$.
The NCSC is described by $H_\text{NCSC}=H_\text{normal}+H_\text{pair}$, where
\begin{equation}
H_\text{normal}=\sum_{\vec k,\sigma,\sigma^{\prime}}f^{\dagger}_{\vec k\sigma}\left(\epsilon^\text{NCSC}_{\vec k}\delta_{\sigma,\sigma^{\prime}}+\boldsymbol{\gamma}_{\vec k}\cdot\boldsymbol{\tau}_{\sigma\sigma^{\prime}}\right)f^{}_{\vec k\sigma^{\prime}}
\end{equation}
and
\begin{equation}
H_\text{pair}=\sum_{\vec k,\sigma,\sigma^{\prime}}(f^{\dagger}_{-\vec k\sigma}\Delta_\text{NCSC}(\vec k)_{\sigma,\sigma^{\prime}}f^{\dagger}_{\vec k\sigma^{\prime}} +\text{H.c.}).
\end{equation}
Here, $\epsilon_{\vec k}^\text{NCSC}$ represents the bare band dispersion. The chemical potential of both superconducting leads is set to zero and chosen as the reference for all other energies. $\boldsymbol{\gamma}_{\vec k}$ represents the antisymmetric spin-orbit coupling, $\boldsymbol{\tau}$ are the Pauli matrices, and 
\begin{equation}
\Delta_\text{NCSC}(\vec k)_{\sigma,\sigma^{\prime}}=\left[(\psi_{\vec k}\mathbb{I} + {\bf d}_{\vec k}\cdot\boldsymbol{\tau})i\tau^y\right]_{\sigma\sigma^\prime},
\end{equation}
is the energy gap matrix in spin space ($\mathbb{I}$ denotes the identity matrix in spin space). Moreover, $\psi_{\vec k}$ and ${\bf d}_{\vec k}$ represent the singlet and the triplet part of the pair potential, respectively. As the Josephson current depends only on the phase difference between the two superconductors, we have chosen a gauge where the order parameters of the NCSC are real.

The Hamiltonian describing the double quantum dot reads
\begin{equation}
	H_\text{DQD}=\sum_i \varepsilon_i n_{i\sigma}+\sum_i \vec B_i\cdot\vec S_i+\sum_i U_in_{i\up}n_{i\down}+U\sum_{\sigma\sigma'}n_{1\sigma}n_{2\sigma'}.
\end{equation}
Here, $\varepsilon_i$ denotes the energy of the single orbital relevant for transport in quantum dot $i=\text{1 and 2}$. In an experiment, the level positions can easily be tuned by gate voltages. The second term describes the magnetic field $\vec B_i$ (measured in units of $g\mu_\text{B}$ with the electron $g$ factor and the Bohr magneton $\mu_\text{B}$) acting on the dot spin $\vec S_i=\frac{\hbar}{2}\sum_{\sigma\sigma'} c_{i\sigma}^\dagger \boldsymbol \sigma_{\sigma\sigma'}c_{i\sigma'}$ with Pauli matrices $\boldsymbol \sigma$. The magnetic fields required for our proposal are of the order of less than \unit[100]{mT} and thus smaller than the critical field of the superconducting electrodes.
The third term describes local Coulomb interactions $U_i$ (the intradot charging energy). In the following, we assume the local Coulomb interactions to be the largest energy scale, $U_i\to\infty$, such that double occupancy of an individual dot is forbidden. This is a good approximation as charging energies of small quantum dots are of the order of several meV and, thus, much larger than the superconducting gaps.
Finally, the last term describes nonlocal Coulomb interactions $U$, which is the energy cost of having each dot occupied with one electron at the same time. For later convenience, we introduce the detuning $\delta=\varepsilon_\text{1}+\varepsilon_\text{2}+U$, which characterizes the deviation from the particle-hole symmetric point $\delta=0$.

The tunnel Hamiltonian,
\begin{equation}
K_\text{BCS}^\text{DQD}=\sum_{\vec k\sigma}\left(t^\text{BCS}_{1}d^{\dagger}_{\vec k\sigma}c^{}_{1\sigma}+t^\text{BCS}_{2}d^{\dagger}_{\vec k\sigma}c^{}_{2\sigma}+\text{H.c.}\right),
\end{equation}
describes the tunnel coupling between the BCS superconducting lead and the quantum dots. $c^\dagger_{1\sigma}$ and $c^\dagger_{2\sigma}$ are the creation operators for electrons in quantum dots 1 and 2, respectively. The tunnel coupling between the NCSC lead and the quantum dots is described by  
\begin{equation}
K_\text{NCSC}^\text{DQD}=\sum_{\vec k\sigma}\left(t^\text{NCSC}_{1}f^{\dagger}_{\vec k\sigma}c^{}_{1\sigma}+t^\text{NCSC}_{2}f^{\dagger}_{\vec k\sigma}c^{}_{2\sigma}+\text{H.c.}\right).
\end{equation}

We include tunneling of electrons from the superconducting leads to the quantum dots up to second order in the tunnel amplitude by performing a Schrieffer-Wolff (SW) type transformation that integrates out the superconducting reservoirs~\cite{nigg_detecting_2015}. For simplicity we assume $t_1^\text{BCS}=t_2^\text{BCS}=t_\text{BCS}$ and $t_1^\text{NCSC}=t_2^\text{NCSC}=t_\text{NCSC}$. The aim of this SW transformation is to derive an effective low-energy Hamiltonian that captures the Cooper-pair splitting processes from both superconducting leadsby performing a unitary transformation $e^S H e^{-S}$ such that $[H_\text{DQD}+H_\text{BCS}+H_\text{NCSC},S]=K^\text{DQD}_\text{BCS}+K^\text{DQD}_\text{NCSC}$. We assume that the quantum dots are Coulomb-blockaded and that, furthermore, there are \textit{no} quasiparticles in either of the superconductors which is a good approximation if the superconducting gaps are much larger than the temperature. This gives rise to an effective dot Hamiltonian~\cite{rozhkov_interacting-impurity_2000,karrasch_supercurrent_2009,
meng_self-consistent_2009,sothmann_probing_2010,eldridge_superconducting_2010} of the form
\begin{equation}
	H_\text{eff}=H_\text{DQD}+H^\text{BCS}_\text{prox}+H^\text{NCCS}_\text{prox}.
\end{equation}
Here,
\begin{equation}
	H^\text{BCS}_\text{prox}=-\frac{e^{i\Phi}\Gamma_{SL}}{2}\left(c_{2\up}^\dagger c_{1\down}^\dagger-c_{2\down}^\dagger c_{1\up}^\dagger\right)+\text{H.c.}
\end{equation}
describes the proximity effect due to the coupling to the conventional $s$-wave superconductor with  tunnel coupling strength $\Gamma_{SL}=2\pi\rho_\text{BCS}|t_\text{BCS}|^2$, where $\rho_\text{BCS}$ is the normal-state 
density of states at the chemical potential of the superconductor. Moreover,
\begin{align}
	H^\text{NCCS}_\text{prox}=-\frac{\Gamma_{SR}}{2}\left(c_{2\up}^\dagger c_{1\down}^\dagger-c_{2\down}^\dagger c_{1\up}^\dagger\right)
	-\frac{\Gamma_{T^+}}{2}c_{2\up}^\dagger c_{1\up}^\dagger\nonumber\\
	-\frac{\Gamma_{T^0}}{2}\left(c_{2\up}^\dagger c_{1\down}^\dagger+c_{2\down}^\dagger c_{1\up}^\dagger\right)
	-\frac{\Gamma_{T^-}}{2}c_{2\down}^\dagger c_{1\down}^\dagger+\text{H.c.}
\end{align}
describes the proximity effect due to the coupling to the NCSC. In addition to tunneling of spin-singlet Cooper pairs characterized by the first term with tunnel coupling strength $\Gamma_{SR}$, there are now three additional terms that describe tunneling of spin-triplet Cooper pairs with different $z$ components of the spin and tunnel coupling strengths: $\Gamma_{T^+}$, $\Gamma_{T^0}$ and $\Gamma_{T^-}$. In general these coupling strengths can be expressed in terms of the original parameters of $H^\text{NCSC}$. For example, in the limit where the orbital energies of the quantum dots $\varepsilon_i$ and the magnetic field strengths ${\bf B}_i$ are much smaller than the superconducting gap,
\begin{widetext}
\begin{eqnarray}
	\Gamma_{SR}&=&\sum_{\vec k}\cos(\theta_{{\bf k}})|t_{NCSC}|^2\left(\frac{v_+(\vec k)u_+(\vec k)+v_+(-{\bf k})u_+(-\vec k)}{E_{+}(\vec k)} + \frac{v_-(\vec k)u_-(\vec k)+v_-(-{\bf k})u_-(-\vec k)}{E_{-}(\vec k)}\right)\nonumber \\
	\Gamma_{T^0}&=&\sum_{\vec k}i\sin(\theta_{{\bf k}})|t_{NCSC}|^2\left(\frac{v_+(\vec k)u_+(\vec k)-v_+(-{\bf k})u_+(-\vec k)}{E_{+}(\vec k)}+\frac{v_-(\vec k)u_-(\vec k)-v_-(-{\bf k})u_-(-\vec k)}{E_{-}(\vec k)}\right)\nonumber \\
	\Gamma_{T^+}&=&\sum_{\vec k}\sqrt{\frac{\gamma_{\vec k,x}^2+\gamma_{\vec k,y}^2}{|\boldsymbol{\gamma}_{\vec k}|^2}}\frac{|t_{NCSC}|^2}{2}
	\left(\frac{v_+(\vec k)u_+(\vec k)-v_+({-\bf k})u_+(-\vec k)}{E_{+}(\vec k)}-\frac{v_-(\vec k)u_-(\vec k)-v_-(-\vec k)u_-(-\vec k)}{E_{-}(\vec k)}\right), \nonumber \\
	\Gamma_{T^-}&=&\sum_{\vec k}\sqrt{\frac{\gamma_{\vec k,x}^2+\gamma_{\vec k,y}^2}{|\boldsymbol{\gamma}_{\vec k}|^2}}e^{2i\theta_{{\bf k}}}\frac{|t_{NCSC}|^2}{2}
	\left(\frac{v_+(-\vec k)u_+(-\vec k)-v_+({\bf k})u_+(\vec k)}{E_{+}(\vec k)}-\frac{v_-(-\vec k)u_-(-\vec k)-v_-(\vec k)u_-(\vec k)}{E_{-}(\vec k)}\right), \nonumber 
\end{eqnarray}
\end{widetext}
where
\begin{eqnarray}
u_{\pm}(\vec k)&=&\sqrt{\frac{1}{2}\left(1+\frac{\xi_\pm(\vec k)}{E_\pm(\vec k)}\right)}, \\
v_{\pm}(\vec k)&=&-\sqrt{\frac{1}{2}\left(1-\frac{\xi_\pm(\vec k)}{E_\pm(\vec k)}\right)}\frac{\Delta_{\pm}(\vec k)}{|\Delta_{\pm}(\vec k)|},
\end{eqnarray}
$\xi_\pm(\vec k)=\epsilon_\vec k^\text{NCSC}\pm|\boldsymbol{\gamma}_{\vec k}|$, $\Delta_\pm(\vec k)=\psi_\vec k\pm |{\bf d}_{\vec k}|$, $E_\pm(\vec k)^2=\xi_\pm(\vec k)^2+\Delta_\pm(\vec k)^2$ and 
$\theta_{{\bf k}}=\tan^{-1}\frac{\gamma_y({\bf k})}{\gamma_x({\bf k})}$ is a phase factor associated with the time-reversal operation~\cite{bauer_non-centrosymmetric_2012}. 
As we are not interested in the momentum dependence of the tunneling of Cooper pairs, below we restrict ourselves to numerical values for the various tunnel coupling strengths. In the following, we assume that the NCSC does not break time-reversal symmetry. In this case, we can always ensure that $\Gamma_{T^+}=\Gamma_{T^-}=0$ by choosing the coordinate system such that $\boldsymbol{\gamma}_{\vec k}\parallel \vec e_z$. We note that while $\Gamma_{SR}$ is real, $\Gamma_{T^0}$ is purely imaginary.

In the case where double occupancy of an individual dot is forbidden, a finite Josephson current arises only for an even occupation of the double dot. Thus, there are only five relevant states of the double dot. A basis of this subspace is spanned by the empty dot 
$\ket{0}$, the singlet $\ket{S}=\frac{1}{\sqrt{2}}\left(c_{2\up}^\dagger c_{1\down}^\dagger-c_{2\down}^\dagger c_{1\up}^\dagger\right)\ket{0}$, and the three Cartesian components of the triplet, $\ket{T^x}=\frac{1}{\sqrt{2}}\left(c_{2\down}^\dagger c_{1\down}^\dagger-c_{2\up}^\dagger c_{1\up}^\dagger\right)\ket{0}$, $\ket{T^y}=\frac{i}{\sqrt{2}}\left(c_{2\down}^\dagger c_{1\down}^\dagger+c_{2\up}^\dagger c_{1\up}^\dagger\right)\ket{0}$, and $\ket{T^z}=\frac{1}{\sqrt{2}}\left(c_{2\up}^\dagger c_{1\down}^\dagger+c_{2\down}^\dagger c_{1\up}^\dagger\right)\ket{0}$. In this basis and with the coordinate system chosen such that $\Gamma_{T^\pm}=0$, the effective dot Hamiltonian takes the form
\begin{widetext}
\begin{equation}\label{eq:HamMatrix}
	H_\text{eff}=
	\left(
	\begin{array}{ccccc}
		0 & -\frac{e^{-i\Phi}\Gamma_{SL}+\Gamma_{SR}}{\sqrt{2}} & 0 & 0 & -\frac{\Gamma_{T^0}^*}{\sqrt{2}} \\
		-\frac{e^{i\Phi}\Gamma_{SL}+\Gamma_{SR}}{\sqrt{2}} & \delta & -\frac{\Delta B_x}{2} & -\frac{\Delta B_y}{2} & -\frac{\Delta B_z}{2} \\
		0 & -\frac{\Delta B_x}{2} & \delta & -i \bar B_z & i \bar B_y \\
		0 & -\frac{\Delta B_y}{2} & i \bar B_z & \delta & -i \bar B_x \\
		-\frac{\Gamma_{T^0}}{\sqrt{2}} & -\frac{\Delta B_z}{2} & -i \bar B_y & i \bar B_x & \delta
	\end{array}
	\right).
\end{equation}
\end{widetext}
From Eq.~\eqref{eq:HamMatrix} we infer that $\Gamma_{SL}$ and $\Gamma_{SR}$ introduce a coupling between the empty dot state and the singlet state. Similarly, $\Gamma_{T^0}$ leads to a coupling between the empty state and the triplet $\ket{T^z}$. Transitions between the singlet and triplet states arise from the difference of magnetic fields, $\vec{\Delta B}=\vec B_1-\vec B_2$, while transitions between the different triplet states are induced by the average magnetic field $\vec {\bar B}=(\vec B_1+\vec B_2)/2$.

The Josephson current through the system follows from the phase dependence of the dot eigenenergies $E_k$ via
\begin{equation}
	I_\text{jos}=\frac{2e}{\hbar}\frac{\partial F}{\partial \Phi}
\end{equation}
where $F=-\kBT \log \sum_k e^{-E_k/(\kBT)}$ denotes the free energy of the quantum dot and $T$ is the temperature of the superconducting leads.

In general, there are no simple analytical expressions for the eigenenergies $E_k$. However, in the absence of any magnetic field, $\vec B_1=\vec B_2=0$, we find 
\begin{subequations}
\label{eq:E0}
\begin{align}
	\label{eq:E0_1}E_\pm&=\frac{\delta}{2}\pm\varepsilon_A,\\
	\label{eq:E0_2}E_{T^\sigma}&=\delta,
\end{align}
\end{subequations}
where $\varepsilon_A=\frac{1}{2}\sqrt{\delta^2+2(\Gamma_{SL}^2+\Gamma_{SR}^2+|\Gamma_{T^0}|^2)+4\Gamma_{SL}\Gamma_{SR}\cos\Phi}$; i.e., the energies of the two eigenstates corresponding to a superposition of the empty dot and the singlet state in general acquire a nontrivial dependence on the phase difference $\Phi$ while the energies of the three triplet states are independent of $\Phi$.
In the presence of an inhomogeneous magnetic field along the $z$ axis, we obtain to first order in $\bar B_z$ and $\Delta B_z$
\begin{widetext}
\begin{subequations}
\label{eq:E1}
\begin{align}
	\label{eq:E1_1}E_\pm&=\frac{\delta}{2}\pm\varepsilon_A\pm\frac{|\Gamma_{T^0}|[\Gamma_{SL}\cos(\Phi-\Phi_{T^0})+\Gamma_{SR}\cos\Phi_{T^0}]}{\varepsilon_A(\delta\mp2\varepsilon_A)}\frac{\Delta B_z}{2},\\
	\label{eq:E1_2}E_{T^0}&=\delta+\frac{2|\Gamma_{T^0}|[\Gamma_{SL}\cos(\Phi-\Phi_{T^0})+\Gamma_{SR}\cos\Phi_{T^0}]}{4\varepsilon_A^2-\delta^2}\Delta B_z,\\
	\label{eq:E1_3}E_{T^\pm}&=\delta\pm \bar B_z,
\end{align}
\end{subequations}
\end{widetext}
where we introduced the phase $\Phi_{T^0}$ of $\Gamma_{T^0}$.
There is now a nontrivial phase dependence for the energies of the three eigenstates that are a superposition of the empty dot, the singlet state and the triplet component $T^z$ while the energies of the triplet components $T^x$ and $T^y$ remain $\Phi$ independent.

Finally, in the case where an inhomogeneous magnetic field is applied in the $x-y$ plane, the eigenenergies to first order in $\vec{\bar B}$ and $\vec{\Delta B}$ read
\begin{widetext}
\begin{subequations}
\label{eq:E2}
\begin{align}
	\label{eq:E2_1}E_\pm&=\frac{\delta}{2}\pm\varepsilon_A,\\
	\label{eq:E2_2}E_{T^\pm}&=\delta\pm\bar B\sqrt{\frac{4\varepsilon_A^2-\delta^2+4|\Gamma_{T^0}|[\Gamma_{SL}\sin(\Phi-\Phi_{T^0})-\Gamma_{SR}\sin\Phi_{T^0}]}{4\varepsilon_A^2-\delta^2}},\\
	\label{eq:E2_3}E_{T^0}&=\delta,
\end{align}
\end{subequations}
\end{widetext}
i.e. we now get a nontrivial phase dependence for the energies of the four eigenstates that correspond to superpositions of the empty dot, the singlet state, and the two triplet components $T^x$ and $T^y$ while the triplet component $T^z$ has a $\Phi$-independent eigenenergy.

\section{\label{sec:results}Results}
In the following, we discuss the Josephson current through the double quantum dot. We first analyze the phase dependence of the dot energies for the case of a conventional superconductor-triplet superconductor junction. We then turn to the phase dependence of the dot energies for a conventional superconductor-NCSC junction. Finally, we discuss the critical currents as probes of the singlet-triplet ratio that are easily accessible in experiment.

\subsection{\label{ssec:S-T}Conventional superconductor-triplet superconductor junction}
We consider a double-dot Josephson junction consisting of a conventional $s$-wave superconductor and a pure triplet superconductor; i.e., in Eq.~\eqref{eq:HamMatrix} we have $\Gamma_{SR}=0$.
If the double dot is subject to a homogeneous magnetic field only, the singlet and triplet correlations induced on the dot by the respective superconducting lead do not couple to each other. As a result, the Josephson current through the system vanishes exactly.
This changes in the presence of an inhomogeneous magnetic field. There are two qualitatively different ways to achieve a singlet-triplet coupling on the double quantum dot.

\begin{figure}
	\includegraphics[width=\columnwidth]{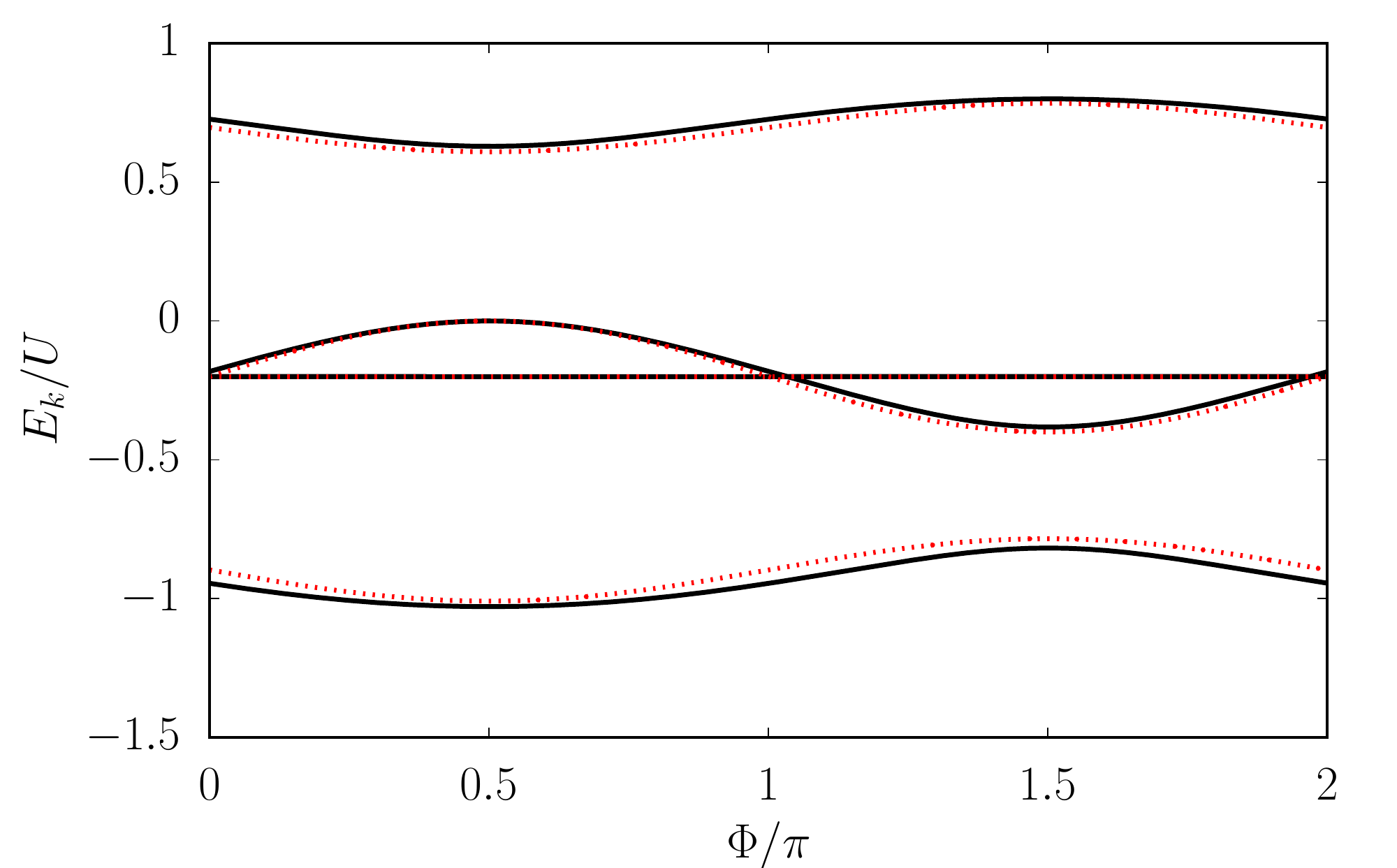}
	\caption{\label{fig:S-T_Bz}(Color online) Phase-dependent eigenenergies of the double quantum dot. Solid lines are the result of numerically diagonalizing Eq.~\eqref{eq:HamMatrix}, dotted lines correspond to the analytical approximation Eqs.~\eqref{eq:E1}. Parameters are $\delta=-0.2U$, $\vec B_1=-\vec B_2=0.25U\vec e_z$, $\Gamma_{SL}=U$, $|\Gamma_{T^0}|=0.5U$, and $\Phi_{T^0}=\pi/2$.}
\end{figure}

In the first case, one applies an inhomogeneous magnetic field along the $z$ axis. As can be seen from Eq.~\eqref{eq:HamMatrix}, this leads to transitions between the singlet and the $T^z$ triplet components. Due to this coupling, three of the eigenenergies develop a nontrivial $\Phi$ dependence [cf. Eqs.~\eqref{eq:E1} as well as Fig.~\ref{fig:S-T_Bz}]. As a result, there is now a finite Josephson current through the double dot. For sufficiently weak values of the inhomogeneity $\Delta B_z$~\footnote{The perturbative expansion in $\Delta B_z$ is valid as long as the energy difference between $\ket{T^Z}$ and $\ket{0}$ as well as $\ket{S}$ is much larger than $\Delta B_z$.}, the Josephson current exhibits a purely sinusoidal dependence $\sin(\Phi-\Phi_{T^0})$ on the phase difference and a magnitude that is proportional to $\Delta B_z$. Importantly, the phase $\Phi_{T^0}=\pi/2$ due to the imaginary value of $\Gamma_{T^0}$ leads to a phase shift of $\pi/2$ in the current-phase relationship.

\begin{figure}
	\includegraphics[width=\columnwidth]{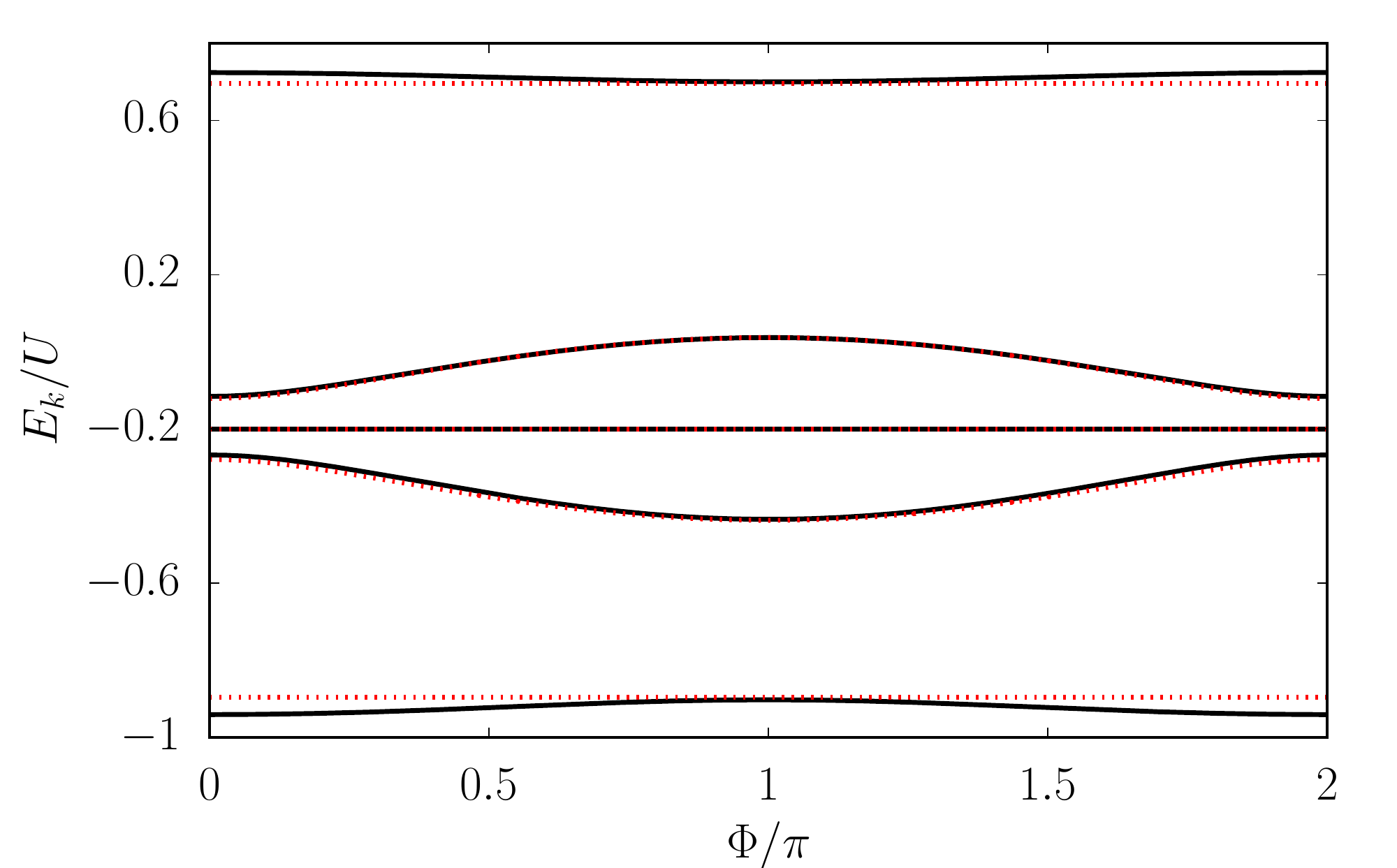}
	\caption{\label{fig:S-T_Bxy}(Color online) Phase-dependent eigenenergies of the double quantum dot. Solid lines are the result of numerically diagonalizing Eq.~\eqref{eq:HamMatrix}, dotted lines correspond to the analytical approximation Eqs.~\eqref{eq:E2}. Parameters are $\delta=-0.2U$, $\vec B_1=0.25U\vec e_x$ $\vec B_2=0.25U\vec e_y$, $\Gamma_{SL}=U$, $|\Gamma_{T^0}|=0.5U$, and $\Phi_{T^0}=\pi/2$.}
\end{figure}

In the second case, one applies an inhomogeneous magnetic field in the $x-y$ plane such that $\bar{\vec B}$ and $\vec{\Delta B}$  are noncollinear. In this situation, the inhomogeneity $\vec{\Delta B}$ converts singlet Cooper pairs into $T^{x,y}$ triplet pairs. 
These triplet pairs are then converted into the $T^z$ triplet component by the average magnetic field $\bar{\vec B}$. As a result there are now four eigenenergies that show a $\Phi$ dependence and thereby give rise to a finite Josephson response [see Eqs.~\eqref{eq:E2} and Fig.~\ref{fig:S-T_Bxy}]. 
From Eq.~\eqref{eq:HamMatrix} we infer that the conversion between the different $S_z$ components of the triplet correlations is accompanied by the phase factor $i$. As a result, the current-phase relationship acquires an additional phaseshift of $\pi/2$ compared to the case where an inhomogeneous magnetic field is applied along the $z$ axis leading to a conventional 0-junction behavior.

\subsection{\label{ssec:S-N}Conventional superconductor-NCSC junction}
\begin{figure}
	\includegraphics[width=\columnwidth]{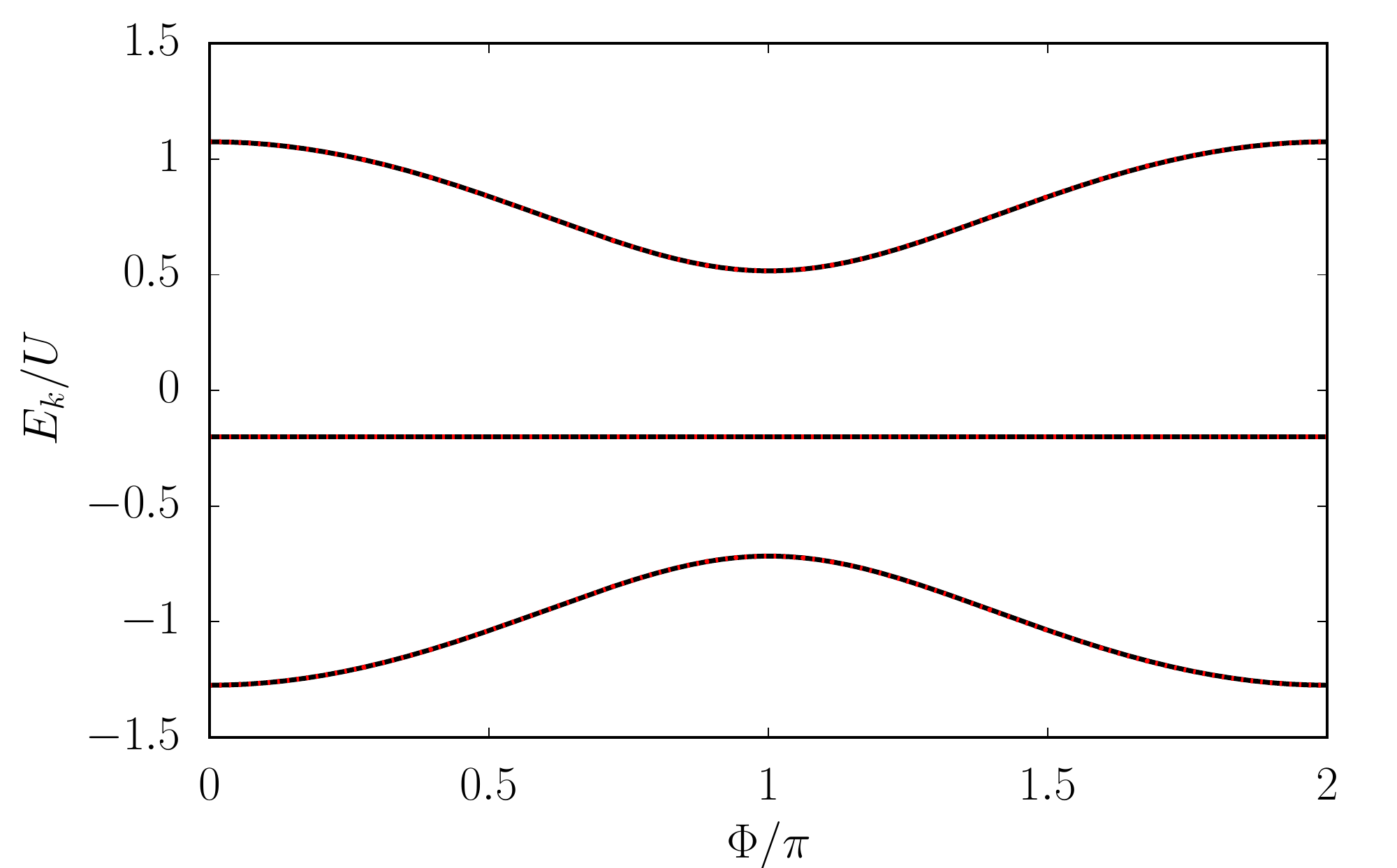}
	\caption{\label{fig:S-ST_B0}(Color online) Phase-dependent eigenenergies of the double quantum dot. Solid lines are the result of numerically diagonalizing Eq.~\eqref{eq:HamMatrix}, dotted lines correspond to the analytical approximation Eqs.~\eqref{eq:E1}. Parameters are $\delta=-0.2U$, $\vec B_1=\vec B_2=0$, $\Gamma_{SL}=U$, $\Gamma_{SR}=0.5U$, $|\Gamma_{T^0}|=0.7U$, and $\Phi_{T^0}=\pi/2$.}
\end{figure}
We now turn to the case of a conventional superconductor-NCSC junction. The most notable difference to the singlet-triplet junction discussed in Sec.~\ref{ssec:S-T} is that now we already get a finite Josephson current even in the absence of any magnetic field [cf. Eqs.~\eqref{eq:E0} and Fig.~\ref{fig:S-ST_B0}]. This is a consequence of the fact that there are now singlet Cooper pairs in \emph{both} superconducting leads that can couple to each other via the double dot. Interestingly, the presence of triplet pairing in the NCSC influences the eigenenergies of the double dot even though the triplet Cooper pairs do not participate in the Josephson response. Thus, in principle one can reconstruct the singlet-triplet ratio from a measurement of the phase-dependent Josephson current in the absence of any magnetic field. 
However, as this requires a precise measurement of the current-phase relationship obtaining the singlet-triplet ratio this way is experimentally very challenging. In the following, we discuss an alternative way to probe the singlet-triplet ratio in the presence of magnetic fields.

\begin{figure}
	\includegraphics[width=\columnwidth]{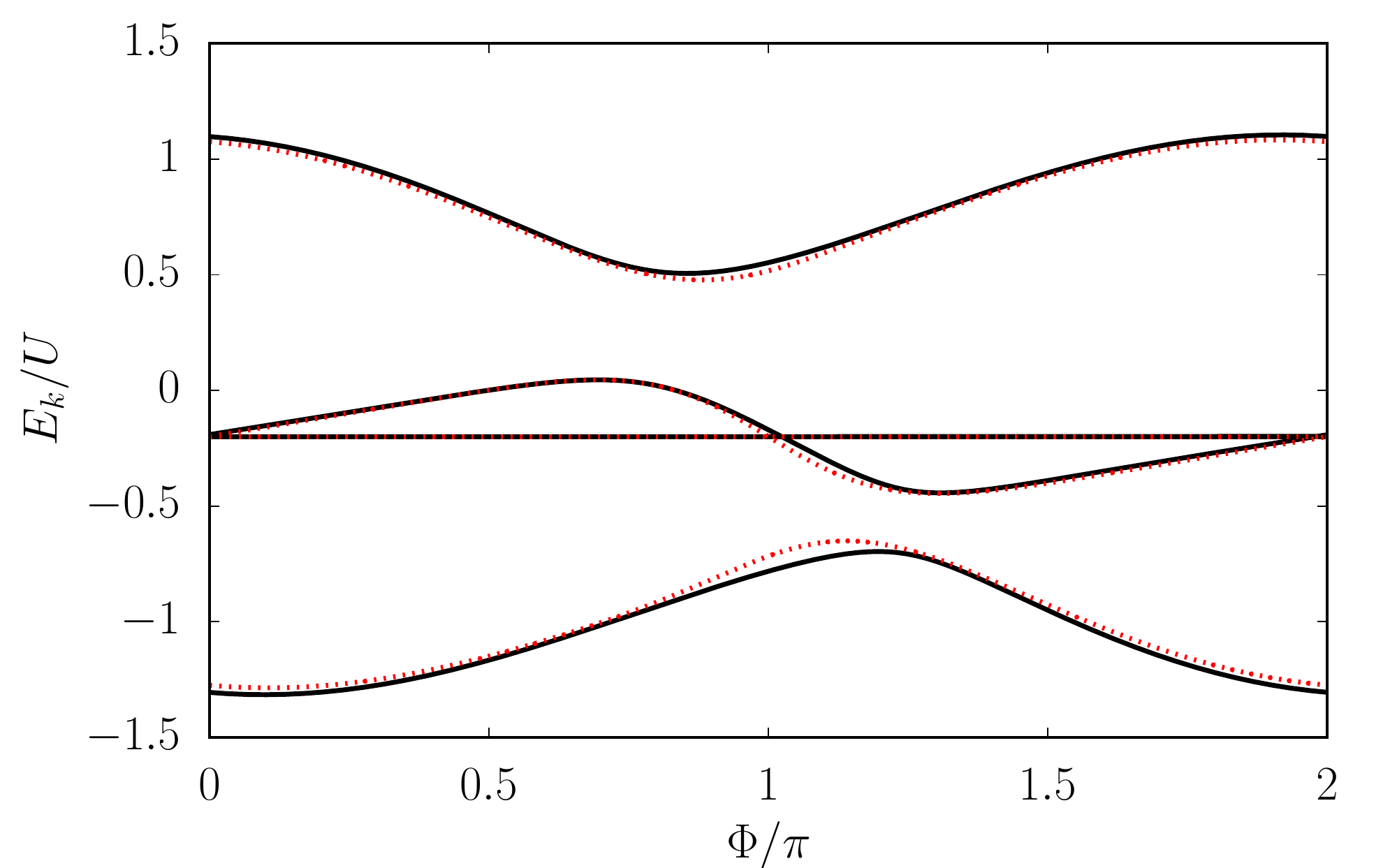}
	\caption{\label{fig:S-ST_Bz}(Color online) Phase-dependent eigenenergies of the double quantum dot. Solid lines are the result of numerically diagonalizing Eq.~\eqref{eq:HamMatrix}, dotted lines correspond to the analytical approximation Eqs.~\eqref{eq:E1}. Parameters are $\delta=-0.2U$, $\vec B_1=-\vec B_2=0.25U\vec e_z$, $\Gamma_{SL}=U$, $\Gamma_{SR}=0.5U$, $|\Gamma_{T^0}|=0.7U$, and $\Phi_{T^0}=\pi/2$.}
\end{figure}
\begin{figure}
	\includegraphics[width=\columnwidth]{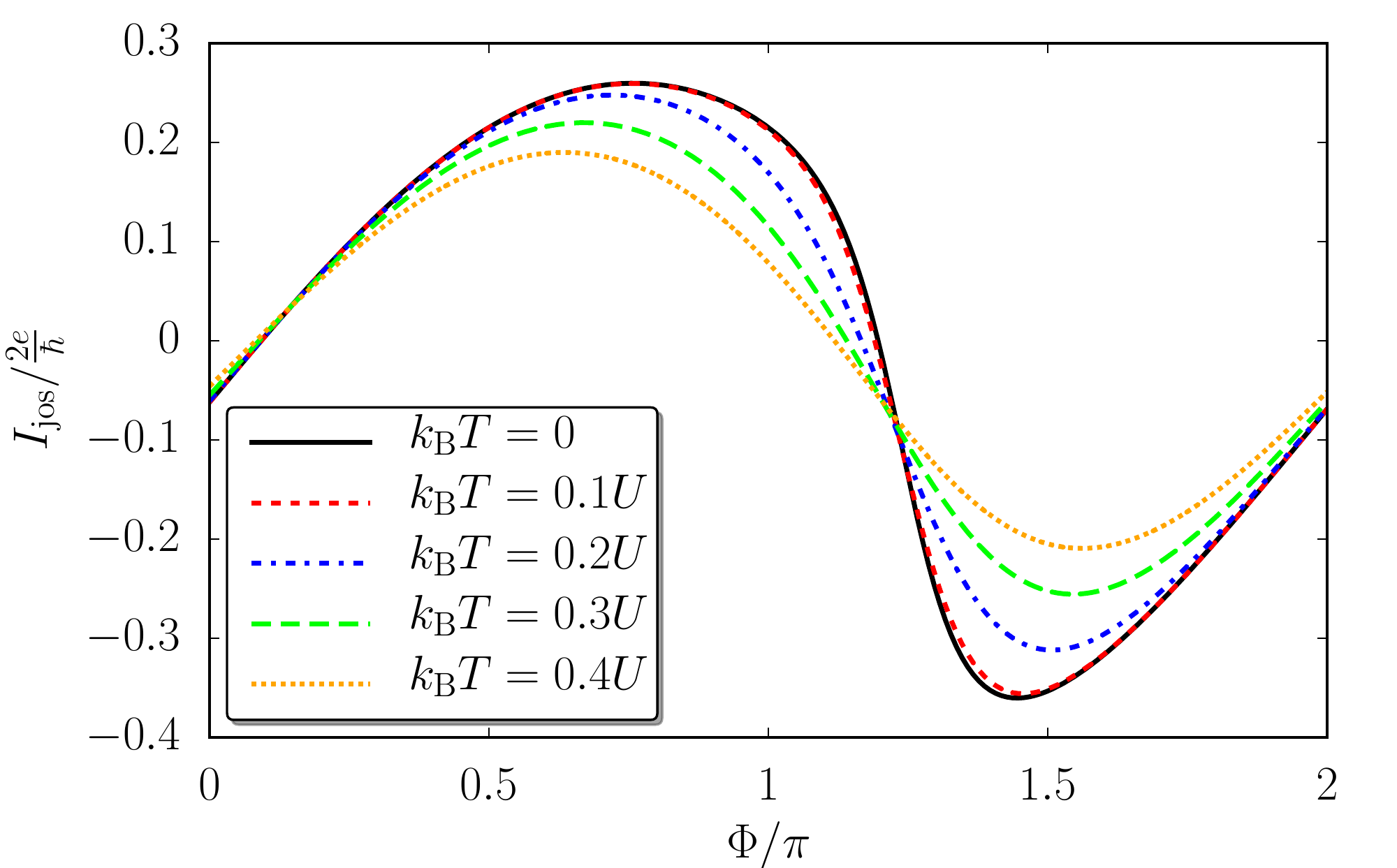}
	\caption{\label{fig:Iphi}(Color online) Current-phase relation for different temperatures. Parameters are chosen as in Fig.~\ref{fig:S-ST_Bz}.}
\end{figure}

We consider the case where an inhomogeneous magnetic field is applied along the $z$ axis, which couples the singlet component of the conventional superconductor to the triplet component $\ket{T^z}$ in the NCSC. As a result, there are again three eigenenergies that develop a nontrivial phase dependence, cf. Fig.~\ref{fig:S-ST_Bz},  similar to the case of the singlet-triplet junction discussed in Sec.~\ref{ssec:S-T}. Interestingly, the contribution to the Josephson current that arises from the singlet and the triplet pairing in the NCSC shows qualitatively different behavior. While the singlet gives rise to a $0$-junction behavior, the triplet component leads to a $\pi/2$-junction behavior. Thus, the interplay between the two components present in a NCSC will generically give rise to a $\varphi$-junction behavior; i.e., there is a finite Josephson current at zero phase bias while there is a vanishing Josephson current at a finite phase difference, cf. Fig.~\ref{fig:Iphi}. It is this nontrivial behavior of the current-phase relation that allows for a precise determination of the singlet-triplet ratio in our setup.

\subsection{Critical current}
As discussed above, the current-phase relation provides detailed insights into the singlet-triplet ratio of a NCSC. Unfortunately, measuring current-phase relations for nanoscale circuits is experimentally very challenging (though the current-phase relation of a small Josephson junction as well as of a carbon nanotube junction was recently detected experimentally~\cite{basset_joint_2014}). Quantities that are routinely measured for mesoscopic Josephson junctions are the critical currents, i.e., the maximal supercurrent that can flow through the device. In the following, we elucidate how this quantity can be used as a probe of the singlet-triplet ratio of a NCSC.

\begin{figure}
	\includegraphics[width=\columnwidth]{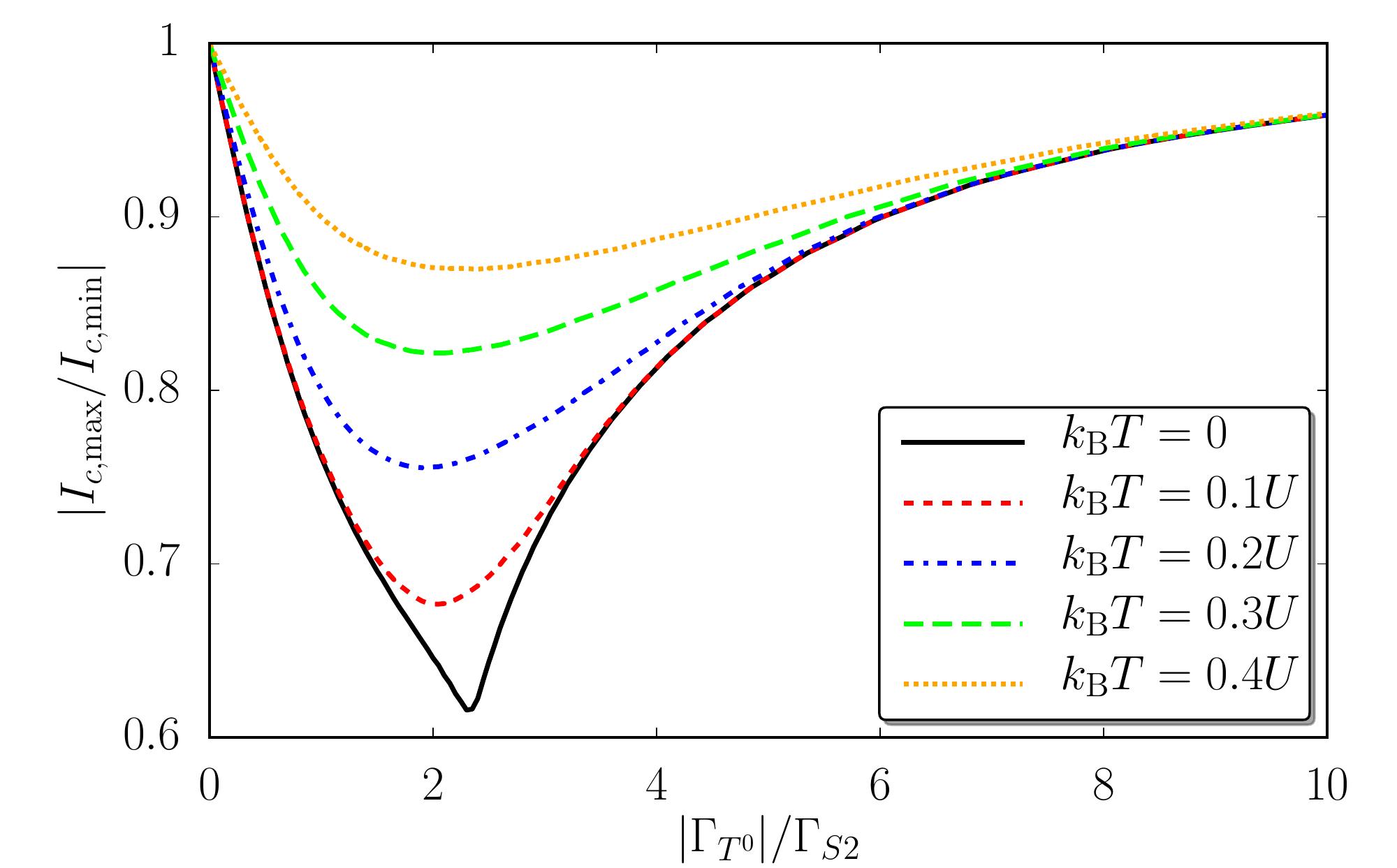}
	\caption{\label{fig:IcRatio}(Color online) Ratio between the maximal and minimal critical current as a function of the triplet-singlet ratio $\Gamma_{T^0}/\Gamma_{SR}$ for different temperatures. Parameters are $\delta=-0.4U$, $\vec B_1=-\vec B_2=0.5U\vec e_z$, $\Gamma_\text{SL}=U$, $\Gamma_\text{SR}=0.4U$, and $\Phi_{T^0}=\pi/2$.}
\end{figure}

An important consequence of the $\varphi$-junction behavior that occurs for any finite singlet-triplet ratio is the fact that the critical current is different for the two directions of current flow [cf. Fig.~\ref{fig:Iphi}]. The breaking of left-right symmetry arises from the interplay of the magnetic fields acting on the double dot and the triplet pairs existing in the NCSC. The difference between the two critical currents (the maximum and minimum value of $I_\text{jos}$) is easily accessible in experiments. In Fig.~\ref{fig:IcRatio} we plot the ratio between the maximal and the minimal critical currents as a function of the singlet-triplet ratio $|\Gamma_{T^0}|/\Gamma_{SR}$. At vanishing $|\Gamma_{T^0}|$ (corresponding to a junction between two conventional $s$-wave superconductors which does not have a $\varphi$-junction behavior) the two currents are equal to each other. We then find that the current ratio decreases with $|\Gamma_{T^0}|$, reaches a minimum and finally goes to unity for $\Gamma_{SR}\to0$ (corresponding to a junction between a conventional $s$-wave superconductor and a triplet superconductor which exhibits a $\pi/2$-junction behavior). Importantly, the precise dependence of the current ratio depends on the temperature of the superconducting leads. Thus, by measuring the ratio between maximal and minimal supercurrents at different temperatures one gets experimental access to the singlet-triplet ratio of a given NCSC material.

We remark that the observation of two different critical currents does not require any fine tuning of parameters. We find that a behavior qualitatively similar to the one shown in Fig.~\ref{fig:IcRatio} occurs 
for any choice of parameters.
However, there is a larger asymmetry for the case of finite detuning $\delta$. Furthermore, for $\Gamma_{SL}>\Gamma_{SR}$ there is in general a stronger asymmetry and the curve has more pronounced features. Most likely having $\Gamma_{SL}>\Gamma_{SR}$ corresponds to the generic scenario of a double-dot junction as discussed here because there are conventional superconductors such as Al that are known to form good contacts to quantum dots~\cite{buitelaar_quantum_2002,buitelaar_multiple_2003,van_dam_supercurrent_2006,buizert_kondo_2007,sand-jespersen_kondo-enhanced_2007,hofstetter_cooper_2009,winkelmann_superconductivity_2009,eichler_tuning_2009,pillet_andreev_2010,katsaros_hybrid_2010,herrmann_carbon_2010,lee_zero-bias_2012} while similar NCSC materials are not known yet.

\section{\label{sec:conclusions}Conclusions}
We theoretically discussed the Josephson current through a double quantum dot coupled to a conventional $s$-wave superconductor as well as a NCSC. We demonstrated that an inhomogeneous magnetic field acting on the double dot can convert singlet Cooper pairs into triplet Cooper pairs and vice versa. This conversion gives rise to interesting features in the current-voltage characteristics such as a $\pi/2$ shift for the triplet component. In particular, we found that for suitably chosen external magnetic fields the junction exhibits a $\varphi$-junction behavior. As a consequence, the critical currents in the forward and backward directions become different. We showed that their ratio measured at different temperatures of the superconducting leads can be used as an accurate probe of the single-triplet ratio in the NCSC.

\acknowledgments
We thank C. Bruder for helpful discussions. We acknowledge financial support from the Swiss National Science Foundation.


\begin{thebibliography}{58}%
\makeatletter
\providecommand \@ifxundefined [1]{%
 \@ifx{#1\undefined}
}%
\providecommand \@ifnum [1]{%
 \ifnum #1\expandafter \@firstoftwo
 \else \expandafter \@secondoftwo
 \fi
}%
\providecommand \@ifx [1]{%
 \ifx #1\expandafter \@firstoftwo
 \else \expandafter \@secondoftwo
 \fi
}%
\providecommand \natexlab [1]{#1}%
\providecommand \enquote  [1]{``#1''}%
\providecommand \bibnamefont  [1]{#1}%
\providecommand \bibfnamefont [1]{#1}%
\providecommand \citenamefont [1]{#1}%
\providecommand \href@noop [0]{\@secondoftwo}%
\providecommand \href [0]{\begingroup \@sanitize@url \@href}%
\providecommand \@href[1]{\@@startlink{#1}\@@href}%
\providecommand \@@href[1]{\endgroup#1\@@endlink}%
\providecommand \@sanitize@url [0]{\catcode `\\12\catcode `\$12\catcode
  `\&12\catcode `\#12\catcode `\^12\catcode `\_12\catcode `\%12\relax}%
\providecommand \@@startlink[1]{}%
\providecommand \@@endlink[0]{}%
\providecommand \url  [0]{\begingroup\@sanitize@url \@url }%
\providecommand \@url [1]{\endgroup\@href {#1}{\urlprefix }}%
\providecommand \urlprefix  [0]{URL }%
\providecommand \Eprint [0]{\href }%
\providecommand \doibase [0]{http://dx.doi.org/}%
\providecommand \selectlanguage [0]{\@gobble}%
\providecommand \bibinfo  [0]{\@secondoftwo}%
\providecommand \bibfield  [0]{\@secondoftwo}%
\providecommand \translation [1]{[#1]}%
\providecommand \BibitemOpen [0]{}%
\providecommand \bibitemStop [0]{}%
\providecommand \bibitemNoStop [0]{.\EOS\space}%
\providecommand \EOS [0]{\spacefactor3000\relax}%
\providecommand \BibitemShut  [1]{\csname bibitem#1\endcsname}%
\let\auto@bib@innerbib\@empty
\bibitem [{\citenamefont {Bardeen}\ \emph {et~al.}(1957)\citenamefont
  {Bardeen}, \citenamefont {Cooper},\ and\ \citenamefont
  {Schrieffer}}]{bardeen_theory_1957}%
  \BibitemOpen
  \bibfield  {author} {\bibinfo {author} {\bibfnamefont {J.}~\bibnamefont
  {Bardeen}}, \bibinfo {author} {\bibfnamefont {L.~N.}\ \bibnamefont {Cooper}},
  \ and\ \bibinfo {author} {\bibfnamefont {J.~R.}\ \bibnamefont {Schrieffer}},\
  }\href {\doibase 10.1103/PhysRev.108.1175} {\bibfield  {journal} {\bibinfo
  {journal} {Phys. Rev.}\ }\textbf {\bibinfo {volume} {108}},\ \bibinfo {pages}
  {1175} (\bibinfo {year} {1957})}\BibitemShut {NoStop}%
\bibitem [{\citenamefont {Bauer}\ \emph {et~al.}(2004)\citenamefont {Bauer},
  \citenamefont {Hilscher}, \citenamefont {Michor}, \citenamefont {Paul},
  \citenamefont {Scheidt}, \citenamefont {Gribanov}, \citenamefont {Seropegin},
  \citenamefont {Noël}, \citenamefont {Sigrist},\ and\ \citenamefont
  {Rogl}}]{bauer_heavy_2004}%
  \BibitemOpen
  \bibfield  {author} {\bibinfo {author} {\bibfnamefont {E.}~\bibnamefont
  {Bauer}}, \bibinfo {author} {\bibfnamefont {G.}~\bibnamefont {Hilscher}},
  \bibinfo {author} {\bibfnamefont {H.}~\bibnamefont {Michor}}, \bibinfo
  {author} {\bibfnamefont {C.}~\bibnamefont {Paul}}, \bibinfo {author}
  {\bibfnamefont {E.~W.}\ \bibnamefont {Scheidt}}, \bibinfo {author}
  {\bibfnamefont {A.}~\bibnamefont {Gribanov}}, \bibinfo {author}
  {\bibfnamefont {Y.}~\bibnamefont {Seropegin}}, \bibinfo {author}
  {\bibfnamefont {H.}~\bibnamefont {Noël}}, \bibinfo {author} {\bibfnamefont
  {M.}~\bibnamefont {Sigrist}}, \ and\ \bibinfo {author} {\bibfnamefont
  {P.}~\bibnamefont {Rogl}},\ }\href {\doibase 10.1103/PhysRevLett.92.027003}
  {\bibfield  {journal} {\bibinfo  {journal} {Phys. Rev. Lett.}\ }\textbf
  {\bibinfo {volume} {92}},\ \bibinfo {pages} {027003} (\bibinfo {year}
  {2004})}\BibitemShut {NoStop}%
\bibitem [{\citenamefont {Kimura}\ \emph {et~al.}(2005)\citenamefont {Kimura},
  \citenamefont {Ito}, \citenamefont {Saitoh}, \citenamefont {Umeda},
  \citenamefont {Aoki},\ and\ \citenamefont
  {Terashima}}]{kimura_pressure-induced_2005}%
  \BibitemOpen
  \bibfield  {author} {\bibinfo {author} {\bibfnamefont {N.}~\bibnamefont
  {Kimura}}, \bibinfo {author} {\bibfnamefont {K.}~\bibnamefont {Ito}},
  \bibinfo {author} {\bibfnamefont {K.}~\bibnamefont {Saitoh}}, \bibinfo
  {author} {\bibfnamefont {Y.}~\bibnamefont {Umeda}}, \bibinfo {author}
  {\bibfnamefont {H.}~\bibnamefont {Aoki}}, \ and\ \bibinfo {author}
  {\bibfnamefont {T.}~\bibnamefont {Terashima}},\ }\href {\doibase
  10.1103/PhysRevLett.95.247004} {\bibfield  {journal} {\bibinfo  {journal}
  {Phys. Rev. Lett.}\ }\textbf {\bibinfo {volume} {95}},\ \bibinfo {pages}
  {247004} (\bibinfo {year} {2005})}\BibitemShut {NoStop}%
\bibitem [{\citenamefont {Sugitani}\ \emph {et~al.}(2006)\citenamefont
  {Sugitani}, \citenamefont {Okuda}, \citenamefont {Shishido}, \citenamefont
  {Yamada}, \citenamefont {Thamizhavel}, \citenamefont {Yamamoto},
  \citenamefont {D.~Matsuda}, \citenamefont {Haga}, \citenamefont {Takeuchi},
  \citenamefont {Settai},\ and\ \citenamefont
  {Onuki}}]{sugitani_pressure-induced_2006}%
  \BibitemOpen
  \bibfield  {author} {\bibinfo {author} {\bibfnamefont {I.}~\bibnamefont
  {Sugitani}}, \bibinfo {author} {\bibfnamefont {Y.}~\bibnamefont {Okuda}},
  \bibinfo {author} {\bibfnamefont {H.}~\bibnamefont {Shishido}}, \bibinfo
  {author} {\bibfnamefont {T.}~\bibnamefont {Yamada}}, \bibinfo {author}
  {\bibfnamefont {A.}~\bibnamefont {Thamizhavel}}, \bibinfo {author}
  {\bibfnamefont {E.}~\bibnamefont {Yamamoto}}, \bibinfo {author}
  {\bibfnamefont {T.}~\bibnamefont {D.~Matsuda}}, \bibinfo {author}
  {\bibfnamefont {Y.}~\bibnamefont {Haga}}, \bibinfo {author} {\bibfnamefont
  {T.}~\bibnamefont {Takeuchi}}, \bibinfo {author} {\bibfnamefont
  {R.}~\bibnamefont {Settai}}, \ and\ \bibinfo {author} {\bibfnamefont
  {Y.}~\bibnamefont {Onuki}},\ }\href {\doibase 10.1143/JPSJ.75.043703}
  {\bibfield  {journal} {\bibinfo  {journal} {J. Phys. Soc. Jpn.}\ }\textbf
  {\bibinfo {volume} {75}},\ \bibinfo {pages} {043703} (\bibinfo {year}
  {2006})}\BibitemShut {NoStop}%
\bibitem [{\citenamefont {Honda}\ \emph {et~al.}(2010)\citenamefont {Honda},
  \citenamefont {Bonalde}, \citenamefont {Shimizu}, \citenamefont {Yoshiuchi},
  \citenamefont {Hirose}, \citenamefont {Nakamura}, \citenamefont {Settai},\
  and\ \citenamefont {Onuki}}]{honda_pressure-induced_2010}%
  \BibitemOpen
  \bibfield  {author} {\bibinfo {author} {\bibfnamefont {F.}~\bibnamefont
  {Honda}}, \bibinfo {author} {\bibfnamefont {I.}~\bibnamefont {Bonalde}},
  \bibinfo {author} {\bibfnamefont {K.}~\bibnamefont {Shimizu}}, \bibinfo
  {author} {\bibfnamefont {S.}~\bibnamefont {Yoshiuchi}}, \bibinfo {author}
  {\bibfnamefont {Y.}~\bibnamefont {Hirose}}, \bibinfo {author} {\bibfnamefont
  {T.}~\bibnamefont {Nakamura}}, \bibinfo {author} {\bibfnamefont
  {R.}~\bibnamefont {Settai}}, \ and\ \bibinfo {author} {\bibfnamefont
  {Y.}~\bibnamefont {Onuki}},\ }\href {\doibase 10.1103/PhysRevB.81.140507}
  {\bibfield  {journal} {\bibinfo  {journal} {Phys. Rev. B}\ }\textbf {\bibinfo
  {volume} {81}},\ \bibinfo {pages} {140507} (\bibinfo {year}
  {2010})}\BibitemShut {NoStop}%
\bibitem [{\citenamefont {Akazawa}\ \emph {et~al.}(2004)\citenamefont
  {Akazawa}, \citenamefont {Hidaka}, \citenamefont {Kotegawa}, \citenamefont
  {C.~Kobayashi}, \citenamefont {Fujiwara}, \citenamefont {Yamamoto},
  \citenamefont {Haga}, \citenamefont {Settai},\ and\ \citenamefont
  {Onuki}}]{akazawa_pressure-induced_2004}%
  \BibitemOpen
  \bibfield  {author} {\bibinfo {author} {\bibfnamefont {T.}~\bibnamefont
  {Akazawa}}, \bibinfo {author} {\bibfnamefont {H.}~\bibnamefont {Hidaka}},
  \bibinfo {author} {\bibfnamefont {H.}~\bibnamefont {Kotegawa}}, \bibinfo
  {author} {\bibfnamefont {T.}~\bibnamefont {C.~Kobayashi}}, \bibinfo {author}
  {\bibfnamefont {T.}~\bibnamefont {Fujiwara}}, \bibinfo {author}
  {\bibfnamefont {E.}~\bibnamefont {Yamamoto}}, \bibinfo {author}
  {\bibfnamefont {Y.}~\bibnamefont {Haga}}, \bibinfo {author} {\bibfnamefont
  {R.}~\bibnamefont {Settai}}, \ and\ \bibinfo {author} {\bibfnamefont
  {Y.}~\bibnamefont {Onuki}},\ }\href {\doibase 10.1143/JPSJ.73.3129}
  {\bibfield  {journal} {\bibinfo  {journal} {J. Phys. Soc. Jpn.}\ }\textbf
  {\bibinfo {volume} {73}},\ \bibinfo {pages} {3129} (\bibinfo {year}
  {2004})}\BibitemShut {NoStop}%
\bibitem [{\citenamefont {Togano}\ \emph {et~al.}(2004)\citenamefont {Togano},
  \citenamefont {Badica}, \citenamefont {Nakamori}, \citenamefont {Orimo},
  \citenamefont {Takeya},\ and\ \citenamefont
  {Hirata}}]{togano_superconductivity_2004}%
  \BibitemOpen
  \bibfield  {author} {\bibinfo {author} {\bibfnamefont {K.}~\bibnamefont
  {Togano}}, \bibinfo {author} {\bibfnamefont {P.}~\bibnamefont {Badica}},
  \bibinfo {author} {\bibfnamefont {Y.}~\bibnamefont {Nakamori}}, \bibinfo
  {author} {\bibfnamefont {S.}~\bibnamefont {Orimo}}, \bibinfo {author}
  {\bibfnamefont {H.}~\bibnamefont {Takeya}}, \ and\ \bibinfo {author}
  {\bibfnamefont {K.}~\bibnamefont {Hirata}},\ }\href {\doibase
  10.1103/PhysRevLett.93.247004} {\bibfield  {journal} {\bibinfo  {journal}
  {Phys. Rev. Lett.}\ }\textbf {\bibinfo {volume} {93}},\ \bibinfo {pages}
  {247004} (\bibinfo {year} {2004})}\BibitemShut {NoStop}%
\bibitem [{\citenamefont {Badica}\ \emph {et~al.}(2005)\citenamefont {Badica},
  \citenamefont {Kondo},\ and\ \citenamefont
  {Togano}}]{badica_superconductivity_2005}%
  \BibitemOpen
  \bibfield  {author} {\bibinfo {author} {\bibfnamefont {P.}~\bibnamefont
  {Badica}}, \bibinfo {author} {\bibfnamefont {T.}~\bibnamefont {Kondo}}, \
  and\ \bibinfo {author} {\bibfnamefont {K.}~\bibnamefont {Togano}},\ }\href
  {\doibase 10.1143/JPSJ.74.1014} {\bibfield  {journal} {\bibinfo  {journal}
  {J. Phys. Soc. Jpn.}\ }\textbf {\bibinfo {volume} {74}},\ \bibinfo {pages}
  {1014} (\bibinfo {year} {2005})}\BibitemShut {NoStop}%
\bibitem [{\citenamefont {Yuan}\ \emph {et~al.}(2006)\citenamefont {Yuan},
  \citenamefont {Agterberg}, \citenamefont {Hayashi}, \citenamefont {Badica},
  \citenamefont {Vandervelde}, \citenamefont {Togano}, \citenamefont
  {Sigrist},\ and\ \citenamefont {Salamon}}]{yuan_$s$-wave_2006}%
  \BibitemOpen
  \bibfield  {author} {\bibinfo {author} {\bibfnamefont {H.~Q.}\ \bibnamefont
  {Yuan}}, \bibinfo {author} {\bibfnamefont {D.~F.}\ \bibnamefont {Agterberg}},
  \bibinfo {author} {\bibfnamefont {N.}~\bibnamefont {Hayashi}}, \bibinfo
  {author} {\bibfnamefont {P.}~\bibnamefont {Badica}}, \bibinfo {author}
  {\bibfnamefont {D.}~\bibnamefont {Vandervelde}}, \bibinfo {author}
  {\bibfnamefont {K.}~\bibnamefont {Togano}}, \bibinfo {author} {\bibfnamefont
  {M.}~\bibnamefont {Sigrist}}, \ and\ \bibinfo {author} {\bibfnamefont
  {M.~B.}\ \bibnamefont {Salamon}},\ }\href {\doibase
  10.1103/PhysRevLett.97.017006} {\bibfield  {journal} {\bibinfo  {journal}
  {Phys. Rev. Lett.}\ }\textbf {\bibinfo {volume} {97}},\ \bibinfo {pages}
  {017006} (\bibinfo {year} {2006})}\BibitemShut {NoStop}%
\bibitem [{\citenamefont {Bauer}\ \emph {et~al.}(2010)\citenamefont {Bauer},
  \citenamefont {Rogl}, \citenamefont {Chen}, \citenamefont {Khan},
  \citenamefont {Michor}, \citenamefont {Hilscher}, \citenamefont {Royanian},
  \citenamefont {Kumagai}, \citenamefont {Li}, \citenamefont {Li},
  \citenamefont {Podloucky},\ and\ \citenamefont
  {Rogl}}]{bauer_unconventional_2010}%
  \BibitemOpen
  \bibfield  {author} {\bibinfo {author} {\bibfnamefont {E.}~\bibnamefont
  {Bauer}}, \bibinfo {author} {\bibfnamefont {G.}~\bibnamefont {Rogl}},
  \bibinfo {author} {\bibfnamefont {X.-Q.}\ \bibnamefont {Chen}}, \bibinfo
  {author} {\bibfnamefont {R.~T.}\ \bibnamefont {Khan}}, \bibinfo {author}
  {\bibfnamefont {H.}~\bibnamefont {Michor}}, \bibinfo {author} {\bibfnamefont
  {G.}~\bibnamefont {Hilscher}}, \bibinfo {author} {\bibfnamefont
  {E.}~\bibnamefont {Royanian}}, \bibinfo {author} {\bibfnamefont
  {K.}~\bibnamefont {Kumagai}}, \bibinfo {author} {\bibfnamefont {D.~Z.}\
  \bibnamefont {Li}}, \bibinfo {author} {\bibfnamefont {Y.~Y.}\ \bibnamefont
  {Li}}, \bibinfo {author} {\bibfnamefont {R.}~\bibnamefont {Podloucky}}, \
  and\ \bibinfo {author} {\bibfnamefont {P.}~\bibnamefont {Rogl}},\ }\href
  {\doibase 10.1103/PhysRevB.82.064511} {\bibfield  {journal} {\bibinfo
  {journal} {Phys. Rev. B}\ }\textbf {\bibinfo {volume} {82}},\ \bibinfo
  {pages} {064511} (\bibinfo {year} {2010})}\BibitemShut {NoStop}%
\bibitem [{\citenamefont {Karki}\ \emph {et~al.}(2010)\citenamefont {Karki},
  \citenamefont {Xiong}, \citenamefont {Vekhter}, \citenamefont {Browne},
  \citenamefont {Adams}, \citenamefont {Young}, \citenamefont {Thomas},
  \citenamefont {Chan}, \citenamefont {Kim},\ and\ \citenamefont
  {Prozorov}}]{karki_structure_2010}%
  \BibitemOpen
  \bibfield  {author} {\bibinfo {author} {\bibfnamefont {A.~B.}\ \bibnamefont
  {Karki}}, \bibinfo {author} {\bibfnamefont {Y.~M.}\ \bibnamefont {Xiong}},
  \bibinfo {author} {\bibfnamefont {I.}~\bibnamefont {Vekhter}}, \bibinfo
  {author} {\bibfnamefont {D.}~\bibnamefont {Browne}}, \bibinfo {author}
  {\bibfnamefont {P.~W.}\ \bibnamefont {Adams}}, \bibinfo {author}
  {\bibfnamefont {D.~P.}\ \bibnamefont {Young}}, \bibinfo {author}
  {\bibfnamefont {K.~R.}\ \bibnamefont {Thomas}}, \bibinfo {author}
  {\bibfnamefont {J.~Y.}\ \bibnamefont {Chan}}, \bibinfo {author}
  {\bibfnamefont {H.}~\bibnamefont {Kim}}, \ and\ \bibinfo {author}
  {\bibfnamefont {R.}~\bibnamefont {Prozorov}},\ }\href {\doibase
  10.1103/PhysRevB.82.064512} {\bibfield  {journal} {\bibinfo  {journal} {Phys.
  Rev. B}\ }\textbf {\bibinfo {volume} {82}},\ \bibinfo {pages} {064512}
  (\bibinfo {year} {2010})}\BibitemShut {NoStop}%
\bibitem [{\citenamefont {Amano}\ \emph {et~al.}(2004)\citenamefont {Amano},
  \citenamefont {Akutagawa}, \citenamefont {Muranaka}, \citenamefont
  {Zenitani},\ and\ \citenamefont {Akimitsu}}]{amano_superconductivity_2004}%
  \BibitemOpen
  \bibfield  {author} {\bibinfo {author} {\bibfnamefont {G.}~\bibnamefont
  {Amano}}, \bibinfo {author} {\bibfnamefont {S.}~\bibnamefont {Akutagawa}},
  \bibinfo {author} {\bibfnamefont {T.}~\bibnamefont {Muranaka}}, \bibinfo
  {author} {\bibfnamefont {Y.}~\bibnamefont {Zenitani}}, \ and\ \bibinfo
  {author} {\bibfnamefont {J.}~\bibnamefont {Akimitsu}},\ }\href {\doibase
  10.1143/JPSJ.73.530} {\bibfield  {journal} {\bibinfo  {journal} {J. Phys.
  Soc. Jpn.}\ }\textbf {\bibinfo {volume} {73}},\ \bibinfo {pages} {530}
  (\bibinfo {year} {2004})}\BibitemShut {NoStop}%
\bibitem [{\citenamefont {Butch}\ \emph {et~al.}(2011)\citenamefont {Butch},
  \citenamefont {Syers}, \citenamefont {Kirshenbaum}, \citenamefont {Hope},\
  and\ \citenamefont {Paglione}}]{butch_superconductivity_2011}%
  \BibitemOpen
  \bibfield  {author} {\bibinfo {author} {\bibfnamefont {N.~P.}\ \bibnamefont
  {Butch}}, \bibinfo {author} {\bibfnamefont {P.}~\bibnamefont {Syers}},
  \bibinfo {author} {\bibfnamefont {K.}~\bibnamefont {Kirshenbaum}}, \bibinfo
  {author} {\bibfnamefont {A.~P.}\ \bibnamefont {Hope}}, \ and\ \bibinfo
  {author} {\bibfnamefont {J.}~\bibnamefont {Paglione}},\ }\href {\doibase
  10.1103/PhysRevB.84.220504} {\bibfield  {journal} {\bibinfo  {journal} {Phys.
  Rev. B}\ }\textbf {\bibinfo {volume} {84}},\ \bibinfo {pages} {220504}
  (\bibinfo {year} {2011})}\BibitemShut {NoStop}%
\bibitem [{\citenamefont {Joshi}\ \emph {et~al.}(2011)\citenamefont {Joshi},
  \citenamefont {Thamizhavel},\ and\ \citenamefont
  {Ramakrishnan}}]{joshi_superconductivity_2011}%
  \BibitemOpen
  \bibfield  {author} {\bibinfo {author} {\bibfnamefont {B.}~\bibnamefont
  {Joshi}}, \bibinfo {author} {\bibfnamefont {A.}~\bibnamefont {Thamizhavel}},
  \ and\ \bibinfo {author} {\bibfnamefont {S.}~\bibnamefont {Ramakrishnan}},\
  }\href {\doibase 10.1103/PhysRevB.84.064518} {\bibfield  {journal} {\bibinfo
  {journal} {Phys. Rev. B}\ }\textbf {\bibinfo {volume} {84}},\ \bibinfo
  {pages} {064518} (\bibinfo {year} {2011})}\BibitemShut {NoStop}%
\bibitem [{\citenamefont {Tafti}\ \emph {et~al.}(2013)\citenamefont {Tafti},
  \citenamefont {Fujii}, \citenamefont {Juneau-Fecteau}, \citenamefont
  {René~de Cotret}, \citenamefont {Doiron-Leyraud}, \citenamefont {Asamitsu},\
  and\ \citenamefont {Taillefer}}]{tafti_superconductivity_2013}%
  \BibitemOpen
  \bibfield  {author} {\bibinfo {author} {\bibfnamefont {F.~F.}\ \bibnamefont
  {Tafti}}, \bibinfo {author} {\bibfnamefont {T.}~\bibnamefont {Fujii}},
  \bibinfo {author} {\bibfnamefont {A.}~\bibnamefont {Juneau-Fecteau}},
  \bibinfo {author} {\bibfnamefont {S.}~\bibnamefont {René~de Cotret}},
  \bibinfo {author} {\bibfnamefont {N.}~\bibnamefont {Doiron-Leyraud}},
  \bibinfo {author} {\bibfnamefont {A.}~\bibnamefont {Asamitsu}}, \ and\
  \bibinfo {author} {\bibfnamefont {L.}~\bibnamefont {Taillefer}},\ }\href
  {\doibase 10.1103/PhysRevB.87.184504} {\bibfield  {journal} {\bibinfo
  {journal} {Phys. Rev. B}\ }\textbf {\bibinfo {volume} {87}},\ \bibinfo
  {pages} {184504} (\bibinfo {year} {2013})}\BibitemShut {NoStop}%
\bibitem [{\citenamefont {Gor'kov}\ and\ \citenamefont
  {Rashba}(2001)}]{gorkov_superconducting_2001}%
  \BibitemOpen
  \bibfield  {author} {\bibinfo {author} {\bibfnamefont {L.~P.}\ \bibnamefont
  {Gor'kov}}\ and\ \bibinfo {author} {\bibfnamefont {E.~I.}\ \bibnamefont
  {Rashba}},\ }\href {\doibase 10.1103/PhysRevLett.87.037004} {\bibfield
  {journal} {\bibinfo  {journal} {Phys. Rev. Lett.}\ }\textbf {\bibinfo
  {volume} {87}},\ \bibinfo {pages} {037004} (\bibinfo {year}
  {2001})}\BibitemShut {NoStop}%
\bibitem [{\citenamefont {Bauer}\ and\ \citenamefont
  {Sigrist}(2012)}]{bauer_non-centrosymmetric_2012}%
  \BibitemOpen
  \bibfield  {author} {\bibinfo {author} {\bibfnamefont {E.}~\bibnamefont
  {Bauer}}\ and\ \bibinfo {author} {\bibfnamefont {M.}~\bibnamefont
  {Sigrist}},\ }\href@noop {} {\emph {\bibinfo {title} {Non-{Centrosymmetric}
  {Superconductors}: {Introduction} and {Overview}}}}\ (\bibinfo  {publisher}
  {Springer},\ \bibinfo {address} {Heidelberg ; New York},\ \bibinfo {year}
  {2012})\BibitemShut {NoStop}%
\bibitem [{\citenamefont {Børkje}\ and\ \citenamefont
  {Sudbø}(2006)}]{borkje_tunneling_2006}%
  \BibitemOpen
  \bibfield  {author} {\bibinfo {author} {\bibfnamefont {K.}~\bibnamefont
  {Børkje}}\ and\ \bibinfo {author} {\bibfnamefont {A.}~\bibnamefont
  {Sudbø}},\ }\href {\doibase 10.1103/PhysRevB.74.054506} {\bibfield
  {journal} {\bibinfo  {journal} {Phys. Rev. B}\ }\textbf {\bibinfo {volume}
  {74}},\ \bibinfo {pages} {054506} (\bibinfo {year} {2006})}\BibitemShut
  {NoStop}%
\bibitem [{\citenamefont {Børkje}(2007)}]{borkje_using_2007}%
  \BibitemOpen
  \bibfield  {author} {\bibinfo {author} {\bibfnamefont {K.}~\bibnamefont
  {Børkje}},\ }\href {\doibase 10.1103/PhysRevB.76.184513} {\bibfield
  {journal} {\bibinfo  {journal} {Phys. Rev. B}\ }\textbf {\bibinfo {volume}
  {76}},\ \bibinfo {pages} {184513} (\bibinfo {year} {2007})}\BibitemShut
  {NoStop}%
\bibitem [{\citenamefont {Iniotakis}\ \emph {et~al.}(2007)\citenamefont
  {Iniotakis}, \citenamefont {Hayashi}, \citenamefont {Sawa}, \citenamefont
  {Yokoyama}, \citenamefont {May}, \citenamefont {Tanaka},\ and\ \citenamefont
  {Sigrist}}]{iniotakis_andreev_2007}%
  \BibitemOpen
  \bibfield  {author} {\bibinfo {author} {\bibfnamefont {C.}~\bibnamefont
  {Iniotakis}}, \bibinfo {author} {\bibfnamefont {N.}~\bibnamefont {Hayashi}},
  \bibinfo {author} {\bibfnamefont {Y.}~\bibnamefont {Sawa}}, \bibinfo {author}
  {\bibfnamefont {T.}~\bibnamefont {Yokoyama}}, \bibinfo {author}
  {\bibfnamefont {U.}~\bibnamefont {May}}, \bibinfo {author} {\bibfnamefont
  {Y.}~\bibnamefont {Tanaka}}, \ and\ \bibinfo {author} {\bibfnamefont
  {M.}~\bibnamefont {Sigrist}},\ }\href {\doibase 10.1103/PhysRevB.76.012501}
  {\bibfield  {journal} {\bibinfo  {journal} {Phys. Rev. B}\ }\textbf {\bibinfo
  {volume} {76}},\ \bibinfo {pages} {012501} (\bibinfo {year}
  {2007})}\BibitemShut {NoStop}%
\bibitem [{\citenamefont {Vorontsov}\ \emph {et~al.}(2008)\citenamefont
  {Vorontsov}, \citenamefont {Vekhter},\ and\ \citenamefont
  {Eschrig}}]{vorontsov_surface_2008}%
  \BibitemOpen
  \bibfield  {author} {\bibinfo {author} {\bibfnamefont {A.~B.}\ \bibnamefont
  {Vorontsov}}, \bibinfo {author} {\bibfnamefont {I.}~\bibnamefont {Vekhter}},
  \ and\ \bibinfo {author} {\bibfnamefont {M.}~\bibnamefont {Eschrig}},\ }\href
  {\doibase 10.1103/PhysRevLett.101.127003} {\bibfield  {journal} {\bibinfo
  {journal} {Phys. Rev. Lett.}\ }\textbf {\bibinfo {volume} {101}},\ \bibinfo
  {pages} {127003} (\bibinfo {year} {2008})}\BibitemShut {NoStop}%
\bibitem [{\citenamefont {Fujimoto}(2009)}]{fujimoto_unambiguous_2009}%
  \BibitemOpen
  \bibfield  {author} {\bibinfo {author} {\bibfnamefont {S.}~\bibnamefont
  {Fujimoto}},\ }\href {\doibase 10.1103/PhysRevB.79.220506} {\bibfield
  {journal} {\bibinfo  {journal} {Phys. Rev. B}\ }\textbf {\bibinfo {volume}
  {79}},\ \bibinfo {pages} {220506} (\bibinfo {year} {2009})}\BibitemShut
  {NoStop}%
\bibitem [{\citenamefont {Asano}\ and\ \citenamefont
  {Yamano}(2011)}]{asano_josephson_2011}%
  \BibitemOpen
  \bibfield  {author} {\bibinfo {author} {\bibfnamefont {Y.}~\bibnamefont
  {Asano}}\ and\ \bibinfo {author} {\bibfnamefont {S.}~\bibnamefont {Yamano}},\
  }\href {\doibase 10.1103/PhysRevB.84.064526} {\bibfield  {journal} {\bibinfo
  {journal} {Phys. Rev. B}\ }\textbf {\bibinfo {volume} {84}},\ \bibinfo
  {pages} {064526} (\bibinfo {year} {2011})}\BibitemShut {NoStop}%
\bibitem [{\citenamefont {Rahnavard}\ \emph {et~al.}(2014)\citenamefont
  {Rahnavard}, \citenamefont {Manske},\ and\ \citenamefont
  {Annunziata}}]{rahnavard_magnetic_2014}%
  \BibitemOpen
  \bibfield  {author} {\bibinfo {author} {\bibfnamefont {Y.}~\bibnamefont
  {Rahnavard}}, \bibinfo {author} {\bibfnamefont {D.}~\bibnamefont {Manske}}, \
  and\ \bibinfo {author} {\bibfnamefont {G.}~\bibnamefont {Annunziata}},\
  }\href {\doibase 10.1103/PhysRevB.89.214501} {\bibfield  {journal} {\bibinfo
  {journal} {Phys. Rev. B}\ }\textbf {\bibinfo {volume} {89}},\ \bibinfo
  {pages} {214501} (\bibinfo {year} {2014})}\BibitemShut {NoStop}%
\bibitem [{\citenamefont {Klam}\ \emph {et~al.}(2014)\citenamefont {Klam},
  \citenamefont {Epp}, \citenamefont {Chen}, \citenamefont {Sigrist},\ and\
  \citenamefont {Manske}}]{klam_josephson_2014}%
  \BibitemOpen
  \bibfield  {author} {\bibinfo {author} {\bibfnamefont {L.}~\bibnamefont
  {Klam}}, \bibinfo {author} {\bibfnamefont {A.}~\bibnamefont {Epp}}, \bibinfo
  {author} {\bibfnamefont {W.}~\bibnamefont {Chen}}, \bibinfo {author}
  {\bibfnamefont {M.}~\bibnamefont {Sigrist}}, \ and\ \bibinfo {author}
  {\bibfnamefont {D.}~\bibnamefont {Manske}},\ }\href {\doibase
  10.1103/PhysRevB.89.174505} {\bibfield  {journal} {\bibinfo  {journal} {Phys.
  Rev. B}\ }\textbf {\bibinfo {volume} {89}},\ \bibinfo {pages} {174505}
  (\bibinfo {year} {2014})}\BibitemShut {NoStop}%
\bibitem [{\citenamefont {Klam}\ \emph {et~al.}(2009)\citenamefont {Klam},
  \citenamefont {Einzel},\ and\ \citenamefont {Manske}}]{klam_electronic_2009}%
  \BibitemOpen
  \bibfield  {author} {\bibinfo {author} {\bibfnamefont {L.}~\bibnamefont
  {Klam}}, \bibinfo {author} {\bibfnamefont {D.}~\bibnamefont {Einzel}}, \ and\
  \bibinfo {author} {\bibfnamefont {D.}~\bibnamefont {Manske}},\ }\href
  {\doibase 10.1103/PhysRevLett.102.027004} {\bibfield  {journal} {\bibinfo
  {journal} {Phys. Rev. Lett.}\ }\textbf {\bibinfo {volume} {102}},\ \bibinfo
  {pages} {027004} (\bibinfo {year} {2009})}\BibitemShut {NoStop}%
\bibitem [{\citenamefont {Choi}\ \emph {et~al.}(2000)\citenamefont {Choi},
  \citenamefont {Bruder},\ and\ \citenamefont
  {Loss}}]{choi_spin-dependent_2000}%
  \BibitemOpen
  \bibfield  {author} {\bibinfo {author} {\bibfnamefont {M.-S.}\ \bibnamefont
  {Choi}}, \bibinfo {author} {\bibfnamefont {C.}~\bibnamefont {Bruder}}, \ and\
  \bibinfo {author} {\bibfnamefont {D.}~\bibnamefont {Loss}},\ }\href {\doibase
  10.1103/PhysRevB.62.13569} {\bibfield  {journal} {\bibinfo  {journal} {Phys.
  Rev. B}\ }\textbf {\bibinfo {volume} {62}},\ \bibinfo {pages} {13569}
  (\bibinfo {year} {2000})}\BibitemShut {NoStop}%
\bibitem [{\citenamefont {Tiwari}\ \emph {et~al.}(2014)\citenamefont {Tiwari},
  \citenamefont {Belzig}, \citenamefont {Sigrist},\ and\ \citenamefont
  {Bruder}}]{tiwari_quantum_2014}%
  \BibitemOpen
  \bibfield  {author} {\bibinfo {author} {\bibfnamefont {R.~P.}\ \bibnamefont
  {Tiwari}}, \bibinfo {author} {\bibfnamefont {W.}~\bibnamefont {Belzig}},
  \bibinfo {author} {\bibfnamefont {M.}~\bibnamefont {Sigrist}}, \ and\
  \bibinfo {author} {\bibfnamefont {C.}~\bibnamefont {Bruder}},\ }\href
  {\doibase 10.1103/PhysRevB.89.184512} {\bibfield  {journal} {\bibinfo
  {journal} {Phys. Rev. B}\ }\textbf {\bibinfo {volume} {89}},\ \bibinfo
  {pages} {184512} (\bibinfo {year} {2014})}\BibitemShut {NoStop}%
\bibitem [{\citenamefont {Sothmann}\ \emph {et~al.}(2014)\citenamefont
  {Sothmann}, \citenamefont {Weiss}, \citenamefont {Governale},\ and\
  \citenamefont {König}}]{sothmann_unconventional_2014}%
  \BibitemOpen
  \bibfield  {author} {\bibinfo {author} {\bibfnamefont {B.}~\bibnamefont
  {Sothmann}}, \bibinfo {author} {\bibfnamefont {S.}~\bibnamefont {Weiss}},
  \bibinfo {author} {\bibfnamefont {M.}~\bibnamefont {Governale}}, \ and\
  \bibinfo {author} {\bibfnamefont {J.}~\bibnamefont {König}},\ }\href
  {\doibase 10.1103/PhysRevB.90.220501} {\bibfield  {journal} {\bibinfo
  {journal} {Phys. Rev. B}\ }\textbf {\bibinfo {volume} {90}},\ \bibinfo
  {pages} {220501} (\bibinfo {year} {2014})}\BibitemShut {NoStop}%
\bibitem [{\citenamefont {Buzdin}(2008)}]{buzdin_direct_2008}%
  \BibitemOpen
  \bibfield  {author} {\bibinfo {author} {\bibfnamefont {A.}~\bibnamefont
  {Buzdin}},\ }\href {\doibase 10.1103/PhysRevLett.101.107005} {\bibfield
  {journal} {\bibinfo  {journal} {Phys. Rev. Lett.}\ }\textbf {\bibinfo
  {volume} {101}},\ \bibinfo {pages} {107005} (\bibinfo {year}
  {2008})}\BibitemShut {NoStop}%
\bibitem [{\citenamefont {Brunetti}\ \emph {et~al.}(2013)\citenamefont
  {Brunetti}, \citenamefont {Zazunov}, \citenamefont {Kundu},\ and\
  \citenamefont {Egger}}]{brunetti_anomalous_2013}%
  \BibitemOpen
  \bibfield  {author} {\bibinfo {author} {\bibfnamefont {A.}~\bibnamefont
  {Brunetti}}, \bibinfo {author} {\bibfnamefont {A.}~\bibnamefont {Zazunov}},
  \bibinfo {author} {\bibfnamefont {A.}~\bibnamefont {Kundu}}, \ and\ \bibinfo
  {author} {\bibfnamefont {R.}~\bibnamefont {Egger}},\ }\href {\doibase
  10.1103/PhysRevB.88.144515} {\bibfield  {journal} {\bibinfo  {journal} {Phys.
  Rev. B}\ }\textbf {\bibinfo {volume} {88}},\ \bibinfo {pages} {144515}
  (\bibinfo {year} {2013})}\BibitemShut {NoStop}%
\bibitem [{\citenamefont {Yokoyama}\ \emph {et~al.}(2014)\citenamefont
  {Yokoyama}, \citenamefont {Eto},\ and\ \citenamefont
  {Nazarov}}]{yokoyama_anomalous_2014}%
  \BibitemOpen
  \bibfield  {author} {\bibinfo {author} {\bibfnamefont {T.}~\bibnamefont
  {Yokoyama}}, \bibinfo {author} {\bibfnamefont {M.}~\bibnamefont {Eto}}, \
  and\ \bibinfo {author} {\bibfnamefont {Y.~V.}\ \bibnamefont {Nazarov}},\
  }\href {\doibase 10.1103/PhysRevB.89.195407} {\bibfield  {journal} {\bibinfo
  {journal} {Phys. Rev. B}\ }\textbf {\bibinfo {volume} {89}},\ \bibinfo
  {pages} {195407} (\bibinfo {year} {2014})}\BibitemShut {NoStop}%
\bibitem [{\citenamefont {Feinberg}\ and\ \citenamefont
  {Balseiro}(2014)}]{feinberg_spontaneous_2014}%
  \BibitemOpen
  \bibfield  {author} {\bibinfo {author} {\bibfnamefont {D.}~\bibnamefont
  {Feinberg}}\ and\ \bibinfo {author} {\bibfnamefont {C.~A.}\ \bibnamefont
  {Balseiro}},\ }\href {\doibase 10.1103/PhysRevB.90.075432} {\bibfield
  {journal} {\bibinfo  {journal} {Phys. Rev. B}\ }\textbf {\bibinfo {volume}
  {90}},\ \bibinfo {pages} {075432} (\bibinfo {year} {2014})}\BibitemShut
  {NoStop}%
\bibitem [{\citenamefont {Buzdin}\ and\ \citenamefont
  {Koshelev}(2003)}]{buzdin_periodic_2003}%
  \BibitemOpen
  \bibfield  {author} {\bibinfo {author} {\bibfnamefont {A.}~\bibnamefont
  {Buzdin}}\ and\ \bibinfo {author} {\bibfnamefont {A.~E.}\ \bibnamefont
  {Koshelev}},\ }\href {\doibase 10.1103/PhysRevB.67.220504} {\bibfield
  {journal} {\bibinfo  {journal} {Phys. Rev. B}\ }\textbf {\bibinfo {volume}
  {67}},\ \bibinfo {pages} {220504} (\bibinfo {year} {2003})}\BibitemShut
  {NoStop}%
\bibitem [{\citenamefont {Pugach}\ \emph {et~al.}(2010)\citenamefont {Pugach},
  \citenamefont {Goldobin}, \citenamefont {Kleiner},\ and\ \citenamefont
  {Koelle}}]{pugach_method_2010}%
  \BibitemOpen
  \bibfield  {author} {\bibinfo {author} {\bibfnamefont {N.~G.}\ \bibnamefont
  {Pugach}}, \bibinfo {author} {\bibfnamefont {E.}~\bibnamefont {Goldobin}},
  \bibinfo {author} {\bibfnamefont {R.}~\bibnamefont {Kleiner}}, \ and\
  \bibinfo {author} {\bibfnamefont {D.}~\bibnamefont {Koelle}},\ }\href
  {\doibase 10.1103/PhysRevB.81.104513} {\bibfield  {journal} {\bibinfo
  {journal} {Phys. Rev. B}\ }\textbf {\bibinfo {volume} {81}},\ \bibinfo
  {pages} {104513} (\bibinfo {year} {2010})}\BibitemShut {NoStop}%
\bibitem [{\citenamefont {Goldobin}\ \emph {et~al.}(2011)\citenamefont
  {Goldobin}, \citenamefont {Koelle}, \citenamefont {Kleiner},\ and\
  \citenamefont {Mints}}]{goldobin_josephson_2011}%
  \BibitemOpen
  \bibfield  {author} {\bibinfo {author} {\bibfnamefont {E.}~\bibnamefont
  {Goldobin}}, \bibinfo {author} {\bibfnamefont {D.}~\bibnamefont {Koelle}},
  \bibinfo {author} {\bibfnamefont {R.}~\bibnamefont {Kleiner}}, \ and\
  \bibinfo {author} {\bibfnamefont {R.~G.}\ \bibnamefont {Mints}},\ }\href
  {\doibase 10.1103/PhysRevLett.107.227001} {\bibfield  {journal} {\bibinfo
  {journal} {Phys. Rev. Lett.}\ }\textbf {\bibinfo {volume} {107}},\ \bibinfo
  {pages} {227001} (\bibinfo {year} {2011})}\BibitemShut {NoStop}%
\bibitem [{\citenamefont {Kulagina}\ and\ \citenamefont
  {Linder}(2014)}]{kulagina_spin_2014}%
  \BibitemOpen
  \bibfield  {author} {\bibinfo {author} {\bibfnamefont {I.}~\bibnamefont
  {Kulagina}}\ and\ \bibinfo {author} {\bibfnamefont {J.}~\bibnamefont
  {Linder}},\ }\href {\doibase 10.1103/PhysRevB.90.054504} {\bibfield
  {journal} {\bibinfo  {journal} {Phys. Rev. B}\ }\textbf {\bibinfo {volume}
  {90}},\ \bibinfo {pages} {054504} (\bibinfo {year} {2014})}\BibitemShut
  {NoStop}%
\bibitem [{\citenamefont {Sickinger}\ \emph {et~al.}(2012)\citenamefont
  {Sickinger}, \citenamefont {Lipman}, \citenamefont {Weides}, \citenamefont
  {Mints}, \citenamefont {Kohlstedt}, \citenamefont {Koelle}, \citenamefont
  {Kleiner},\ and\ \citenamefont {Goldobin}}]{sickinger_experimental_2012}%
  \BibitemOpen
  \bibfield  {author} {\bibinfo {author} {\bibfnamefont {H.}~\bibnamefont
  {Sickinger}}, \bibinfo {author} {\bibfnamefont {A.}~\bibnamefont {Lipman}},
  \bibinfo {author} {\bibfnamefont {M.}~\bibnamefont {Weides}}, \bibinfo
  {author} {\bibfnamefont {R.~G.}\ \bibnamefont {Mints}}, \bibinfo {author}
  {\bibfnamefont {H.}~\bibnamefont {Kohlstedt}}, \bibinfo {author}
  {\bibfnamefont {D.}~\bibnamefont {Koelle}}, \bibinfo {author} {\bibfnamefont
  {R.}~\bibnamefont {Kleiner}}, \ and\ \bibinfo {author} {\bibfnamefont
  {E.}~\bibnamefont {Goldobin}},\ }\href {\doibase
  10.1103/PhysRevLett.109.107002} {\bibfield  {journal} {\bibinfo  {journal}
  {Phys. Rev. Lett.}\ }\textbf {\bibinfo {volume} {109}},\ \bibinfo {pages}
  {107002} (\bibinfo {year} {2012})}\BibitemShut {NoStop}%
\bibitem [{\citenamefont {Nigg}\ \emph {et~al.}(2015)\citenamefont {Nigg},
  \citenamefont {Tiwari}, \citenamefont {Walter},\ and\ \citenamefont
  {Schmidt}}]{nigg_detecting_2015}%
  \BibitemOpen
  \bibfield  {author} {\bibinfo {author} {\bibfnamefont {S.~E.}\ \bibnamefont
  {Nigg}}, \bibinfo {author} {\bibfnamefont {R.~P.}\ \bibnamefont {Tiwari}},
  \bibinfo {author} {\bibfnamefont {S.}~\bibnamefont {Walter}}, \ and\ \bibinfo
  {author} {\bibfnamefont {T.~L.}\ \bibnamefont {Schmidt}},\ }\href {\doibase
  10.1103/PhysRevB.91.094516} {\bibfield  {journal} {\bibinfo  {journal} {Phys.
  Rev. B}\ }\textbf {\bibinfo {volume} {91}},\ \bibinfo {pages} {094516}
  (\bibinfo {year} {2015})}\BibitemShut {NoStop}%
\bibitem [{\citenamefont {Rozhkov}\ and\ \citenamefont
  {Arovas}(2000)}]{rozhkov_interacting-impurity_2000}%
  \BibitemOpen
  \bibfield  {author} {\bibinfo {author} {\bibfnamefont {A.~V.}\ \bibnamefont
  {Rozhkov}}\ and\ \bibinfo {author} {\bibfnamefont {D.~P.}\ \bibnamefont
  {Arovas}},\ }\href {\doibase 10.1103/PhysRevB.62.6687} {\bibfield  {journal}
  {\bibinfo  {journal} {Phys. Rev. B}\ }\textbf {\bibinfo {volume} {62}},\
  \bibinfo {pages} {6687} (\bibinfo {year} {2000})}\BibitemShut {NoStop}%
\bibitem [{\citenamefont {Karrasch}\ and\ \citenamefont
  {Meden}(2009)}]{karrasch_supercurrent_2009}%
  \BibitemOpen
  \bibfield  {author} {\bibinfo {author} {\bibfnamefont {C.}~\bibnamefont
  {Karrasch}}\ and\ \bibinfo {author} {\bibfnamefont {V.}~\bibnamefont
  {Meden}},\ }\href {\doibase 10.1103/PhysRevB.79.045110} {\bibfield  {journal}
  {\bibinfo  {journal} {Phys. Rev. B}\ }\textbf {\bibinfo {volume} {79}},\
  \bibinfo {pages} {045110} (\bibinfo {year} {2009})}\BibitemShut {NoStop}%
\bibitem [{\citenamefont {Meng}\ \emph {et~al.}(2009)\citenamefont {Meng},
  \citenamefont {Florens},\ and\ \citenamefont
  {Simon}}]{meng_self-consistent_2009}%
  \BibitemOpen
  \bibfield  {author} {\bibinfo {author} {\bibfnamefont {T.}~\bibnamefont
  {Meng}}, \bibinfo {author} {\bibfnamefont {S.}~\bibnamefont {Florens}}, \
  and\ \bibinfo {author} {\bibfnamefont {P.}~\bibnamefont {Simon}},\ }\href
  {\doibase 10.1103/PhysRevB.79.224521} {\bibfield  {journal} {\bibinfo
  {journal} {Phys. Rev. B}\ }\textbf {\bibinfo {volume} {79}},\ \bibinfo
  {pages} {224521} (\bibinfo {year} {2009})}\BibitemShut {NoStop}%
\bibitem [{\citenamefont {Sothmann}\ \emph {et~al.}(2010)\citenamefont
  {Sothmann}, \citenamefont {Futterer}, \citenamefont {Governale},\ and\
  \citenamefont {König}}]{sothmann_probing_2010}%
  \BibitemOpen
  \bibfield  {author} {\bibinfo {author} {\bibfnamefont {B.}~\bibnamefont
  {Sothmann}}, \bibinfo {author} {\bibfnamefont {D.}~\bibnamefont {Futterer}},
  \bibinfo {author} {\bibfnamefont {M.}~\bibnamefont {Governale}}, \ and\
  \bibinfo {author} {\bibfnamefont {J.}~\bibnamefont {König}},\ }\href
  {\doibase 10.1103/PhysRevB.82.094514} {\bibfield  {journal} {\bibinfo
  {journal} {Phys. Rev. B}\ }\textbf {\bibinfo {volume} {82}},\ \bibinfo
  {pages} {094514} (\bibinfo {year} {2010})}\BibitemShut {NoStop}%
\bibitem [{\citenamefont {Eldridge}\ \emph {et~al.}(2010)\citenamefont
  {Eldridge}, \citenamefont {Pala}, \citenamefont {Governale},\ and\
  \citenamefont {König}}]{eldridge_superconducting_2010}%
  \BibitemOpen
  \bibfield  {author} {\bibinfo {author} {\bibfnamefont {J.}~\bibnamefont
  {Eldridge}}, \bibinfo {author} {\bibfnamefont {M.~G.}\ \bibnamefont {Pala}},
  \bibinfo {author} {\bibfnamefont {M.}~\bibnamefont {Governale}}, \ and\
  \bibinfo {author} {\bibfnamefont {J.}~\bibnamefont {König}},\ }\href
  {\doibase 10.1103/PhysRevB.82.184507} {\bibfield  {journal} {\bibinfo
  {journal} {Phys. Rev. B}\ }\textbf {\bibinfo {volume} {82}},\ \bibinfo
  {pages} {184507} (\bibinfo {year} {2010})}\BibitemShut {NoStop}%
\bibitem [{Note1()}]{Note1}%
  \BibitemOpen
  \bibinfo {note} {The perturbative expansion in $\Delta B_z$ is valid as long as the energy difference between $\ket{T^z}$ and $\ket{0}$ as well as $\ket{S}$ is much larger than $\Delta B_z$.}\BibitemShut {NoStop}%
\bibitem [{\citenamefont {Basset}\ \emph {et~al.}(2014)\citenamefont {Basset},
  \citenamefont {Delagrange}, \citenamefont {Weil}, \citenamefont {Kasumov},
  \citenamefont {Bouchiat},\ and\ \citenamefont {Deblock}}]{basset_joint_2014}%
  \BibitemOpen
  \bibfield  {author} {\bibinfo {author} {\bibfnamefont {J.}~\bibnamefont
  {Basset}}, \bibinfo {author} {\bibfnamefont {R.}~\bibnamefont {Delagrange}},
  \bibinfo {author} {\bibfnamefont {R.}~\bibnamefont {Weil}}, \bibinfo {author}
  {\bibfnamefont {A.}~\bibnamefont {Kasumov}}, \bibinfo {author} {\bibfnamefont
  {H.}~\bibnamefont {Bouchiat}}, \ and\ \bibinfo {author} {\bibfnamefont
  {R.}~\bibnamefont {Deblock}},\ }\href {\doibase 10.1063/1.4887354} {\bibfield
   {journal} {\bibinfo  {journal} {J. Appl. Phys.}\ }\textbf {\bibinfo {volume}
  {116}},\ \bibinfo {pages} {024311} (\bibinfo {year} {2014})}\BibitemShut
  {NoStop}%
\bibitem [{\citenamefont {Buitelaar}\ \emph {et~al.}(2002)\citenamefont
  {Buitelaar}, \citenamefont {Nussbaumer},\ and\ \citenamefont
  {Schönenberger}}]{buitelaar_quantum_2002}%
  \BibitemOpen
  \bibfield  {author} {\bibinfo {author} {\bibfnamefont {M.~R.}\ \bibnamefont
  {Buitelaar}}, \bibinfo {author} {\bibfnamefont {T.}~\bibnamefont
  {Nussbaumer}}, \ and\ \bibinfo {author} {\bibfnamefont {C.}~\bibnamefont
  {Schönenberger}},\ }\href {\doibase 10.1103/PhysRevLett.89.256801}
  {\bibfield  {journal} {\bibinfo  {journal} {Phys. Rev. Lett.}\ }\textbf
  {\bibinfo {volume} {89}},\ \bibinfo {pages} {256801} (\bibinfo {year}
  {2002})}\BibitemShut {NoStop}%
\bibitem [{\citenamefont {Buitelaar}\ \emph {et~al.}(2003)\citenamefont
  {Buitelaar}, \citenamefont {Belzig}, \citenamefont {Nussbaumer},
  \citenamefont {Babicacute}, \citenamefont {Bruder},\ and\ \citenamefont
  {Schönenberger}}]{buitelaar_multiple_2003}%
  \BibitemOpen
  \bibfield  {author} {\bibinfo {author} {\bibfnamefont {M.~R.}\ \bibnamefont
  {Buitelaar}}, \bibinfo {author} {\bibfnamefont {W.}~\bibnamefont {Belzig}},
  \bibinfo {author} {\bibfnamefont {T.}~\bibnamefont {Nussbaumer}}, \bibinfo
  {author} {\bibfnamefont {B.}~\bibnamefont {Babicacute}}, \bibinfo {author}
  {\bibfnamefont {C.}~\bibnamefont {Bruder}}, \ and\ \bibinfo {author}
  {\bibfnamefont {C.}~\bibnamefont {Schönenberger}},\ }\href {\doibase
  10.1103/PhysRevLett.91.057005} {\bibfield  {journal} {\bibinfo  {journal}
  {Phys. Rev. Lett.}\ }\textbf {\bibinfo {volume} {91}},\ \bibinfo {pages}
  {057005} (\bibinfo {year} {2003})}\BibitemShut {NoStop}%
\bibitem [{\citenamefont {van Dam}\ \emph {et~al.}(2006)\citenamefont {van
  Dam}, \citenamefont {Nazarov}, \citenamefont {Bakkers}, \citenamefont
  {De~Franceschi},\ and\ \citenamefont
  {Kouwenhoven}}]{van_dam_supercurrent_2006}%
  \BibitemOpen
  \bibfield  {author} {\bibinfo {author} {\bibfnamefont {J.~A.}\ \bibnamefont
  {van Dam}}, \bibinfo {author} {\bibfnamefont {Y.~V.}\ \bibnamefont
  {Nazarov}}, \bibinfo {author} {\bibfnamefont {E.~P. A.~M.}\ \bibnamefont
  {Bakkers}}, \bibinfo {author} {\bibfnamefont {S.}~\bibnamefont
  {De~Franceschi}}, \ and\ \bibinfo {author} {\bibfnamefont {L.~P.}\
  \bibnamefont {Kouwenhoven}},\ }\href {\doibase 10.1038/nature05018}
  {\bibfield  {journal} {\bibinfo  {journal} {Nature}\ }\textbf {\bibinfo
  {volume} {442}},\ \bibinfo {pages} {667} (\bibinfo {year}
  {2006})}\BibitemShut {NoStop}%
\bibitem [{\citenamefont {Buizert}\ \emph {et~al.}(2007)\citenamefont
  {Buizert}, \citenamefont {Oiwa}, \citenamefont {Shibata}, \citenamefont
  {Hirakawa},\ and\ \citenamefont {Tarucha}}]{buizert_kondo_2007}%
  \BibitemOpen
  \bibfield  {author} {\bibinfo {author} {\bibfnamefont {C.}~\bibnamefont
  {Buizert}}, \bibinfo {author} {\bibfnamefont {A.}~\bibnamefont {Oiwa}},
  \bibinfo {author} {\bibfnamefont {K.}~\bibnamefont {Shibata}}, \bibinfo
  {author} {\bibfnamefont {K.}~\bibnamefont {Hirakawa}}, \ and\ \bibinfo
  {author} {\bibfnamefont {S.}~\bibnamefont {Tarucha}},\ }\href {\doibase
  10.1103/PhysRevLett.99.136806} {\bibfield  {journal} {\bibinfo  {journal}
  {Phys. Rev. Lett.}\ }\textbf {\bibinfo {volume} {99}},\ \bibinfo {pages}
  {136806} (\bibinfo {year} {2007})}\BibitemShut {NoStop}%
\bibitem [{\citenamefont {Sand-Jespersen}\ \emph {et~al.}(2007)\citenamefont
  {Sand-Jespersen}, \citenamefont {Paaske}, \citenamefont {Andersen},
  \citenamefont {Grove-Rasmussen}, \citenamefont {Jørgensen}, \citenamefont
  {Aagesen}, \citenamefont {Sørensen}, \citenamefont {Lindelof}, \citenamefont
  {Flensberg},\ and\ \citenamefont
  {Nygård}}]{sand-jespersen_kondo-enhanced_2007}%
  \BibitemOpen
  \bibfield  {author} {\bibinfo {author} {\bibfnamefont {T.}~\bibnamefont
  {Sand-Jespersen}}, \bibinfo {author} {\bibfnamefont {J.}~\bibnamefont
  {Paaske}}, \bibinfo {author} {\bibfnamefont {B.~M.}\ \bibnamefont
  {Andersen}}, \bibinfo {author} {\bibfnamefont {K.}~\bibnamefont
  {Grove-Rasmussen}}, \bibinfo {author} {\bibfnamefont {H.~I.}\ \bibnamefont
  {Jørgensen}}, \bibinfo {author} {\bibfnamefont {M.}~\bibnamefont {Aagesen}},
  \bibinfo {author} {\bibfnamefont {C.~B.}\ \bibnamefont {Sørensen}}, \bibinfo
  {author} {\bibfnamefont {P.~E.}\ \bibnamefont {Lindelof}}, \bibinfo {author}
  {\bibfnamefont {K.}~\bibnamefont {Flensberg}}, \ and\ \bibinfo {author}
  {\bibfnamefont {J.}~\bibnamefont {Nygård}},\ }\href {\doibase
  10.1103/PhysRevLett.99.126603} {\bibfield  {journal} {\bibinfo  {journal}
  {Phys. Rev. Lett.}\ }\textbf {\bibinfo {volume} {99}},\ \bibinfo {pages}
  {126603} (\bibinfo {year} {2007})}\BibitemShut {NoStop}%
\bibitem [{\citenamefont {Hofstetter}\ \emph {et~al.}(2009)\citenamefont
  {Hofstetter}, \citenamefont {Csonka}, \citenamefont {Nygard},\ and\
  \citenamefont {Schönenberger}}]{hofstetter_cooper_2009}%
  \BibitemOpen
  \bibfield  {author} {\bibinfo {author} {\bibfnamefont {L.}~\bibnamefont
  {Hofstetter}}, \bibinfo {author} {\bibfnamefont {S.}~\bibnamefont {Csonka}},
  \bibinfo {author} {\bibfnamefont {J.}~\bibnamefont {Nygard}}, \ and\ \bibinfo
  {author} {\bibfnamefont {C.}~\bibnamefont {Schönenberger}},\ }\href
  {\doibase 10.1038/nature08432} {\bibfield  {journal} {\bibinfo  {journal}
  {Nature}\ }\textbf {\bibinfo {volume} {461}},\ \bibinfo {pages} {960}
  (\bibinfo {year} {2009})}\BibitemShut {NoStop}%
\bibitem [{\citenamefont {Winkelmann}\ \emph {et~al.}(2009)\citenamefont
  {Winkelmann}, \citenamefont {Roch}, \citenamefont {Wernsdorfer},
  \citenamefont {Bouchiat},\ and\ \citenamefont
  {Balestro}}]{winkelmann_superconductivity_2009}%
  \BibitemOpen
  \bibfield  {author} {\bibinfo {author} {\bibfnamefont {C.~B.}\ \bibnamefont
  {Winkelmann}}, \bibinfo {author} {\bibfnamefont {N.}~\bibnamefont {Roch}},
  \bibinfo {author} {\bibfnamefont {W.}~\bibnamefont {Wernsdorfer}}, \bibinfo
  {author} {\bibfnamefont {V.}~\bibnamefont {Bouchiat}}, \ and\ \bibinfo
  {author} {\bibfnamefont {F.}~\bibnamefont {Balestro}},\ }\href {\doibase
  10.1038/nphys1433} {\bibfield  {journal} {\bibinfo  {journal} {Nat. Phys.}\
  }\textbf {\bibinfo {volume} {5}},\ \bibinfo {pages} {876} (\bibinfo {year}
  {2009})}\BibitemShut {NoStop}%
\bibitem [{\citenamefont {Eichler}\ \emph {et~al.}(2009)\citenamefont
  {Eichler}, \citenamefont {Deblock}, \citenamefont {Weiss}, \citenamefont
  {Karrasch}, \citenamefont {Meden}, \citenamefont {Schönenberger},\ and\
  \citenamefont {Bouchiat}}]{eichler_tuning_2009}%
  \BibitemOpen
  \bibfield  {author} {\bibinfo {author} {\bibfnamefont {A.}~\bibnamefont
  {Eichler}}, \bibinfo {author} {\bibfnamefont {R.}~\bibnamefont {Deblock}},
  \bibinfo {author} {\bibfnamefont {M.}~\bibnamefont {Weiss}}, \bibinfo
  {author} {\bibfnamefont {C.}~\bibnamefont {Karrasch}}, \bibinfo {author}
  {\bibfnamefont {V.}~\bibnamefont {Meden}}, \bibinfo {author} {\bibfnamefont
  {C.}~\bibnamefont {Schönenberger}}, \ and\ \bibinfo {author} {\bibfnamefont
  {H.}~\bibnamefont {Bouchiat}},\ }\href {\doibase 10.1103/PhysRevB.79.161407}
  {\bibfield  {journal} {\bibinfo  {journal} {Phys. Rev. B}\ }\textbf {\bibinfo
  {volume} {79}},\ \bibinfo {pages} {161407} (\bibinfo {year}
  {2009})}\BibitemShut {NoStop}%
\bibitem [{\citenamefont {Pillet}\ \emph {et~al.}(2010)\citenamefont {Pillet},
  \citenamefont {Quay}, \citenamefont {Morfin}, \citenamefont {Bena},
  \citenamefont {Yeyati},\ and\ \citenamefont {Joyez}}]{pillet_andreev_2010}%
  \BibitemOpen
  \bibfield  {author} {\bibinfo {author} {\bibfnamefont {J.-D.}\ \bibnamefont
  {Pillet}}, \bibinfo {author} {\bibfnamefont {C.~H.~L.}\ \bibnamefont {Quay}},
  \bibinfo {author} {\bibfnamefont {P.}~\bibnamefont {Morfin}}, \bibinfo
  {author} {\bibfnamefont {C.}~\bibnamefont {Bena}}, \bibinfo {author}
  {\bibfnamefont {A.~L.}\ \bibnamefont {Yeyati}}, \ and\ \bibinfo {author}
  {\bibfnamefont {P.}~\bibnamefont {Joyez}},\ }\href {\doibase
  10.1038/nphys1811} {\bibfield  {journal} {\bibinfo  {journal} {Nat. Phys.}\
  }\textbf {\bibinfo {volume} {6}},\ \bibinfo {pages} {965} (\bibinfo {year}
  {2010})}\BibitemShut {NoStop}%
\bibitem [{\citenamefont {Katsaros}\ \emph {et~al.}(2010)\citenamefont
  {Katsaros}, \citenamefont {Spathis}, \citenamefont {Stoffel}, \citenamefont
  {Fournel}, \citenamefont {Mongillo}, \citenamefont {Bouchiat}, \citenamefont
  {Lefloch}, \citenamefont {Rastelli}, \citenamefont {Schmidt},\ and\
  \citenamefont {Franceschi}}]{katsaros_hybrid_2010}%
  \BibitemOpen
  \bibfield  {author} {\bibinfo {author} {\bibfnamefont {G.}~\bibnamefont
  {Katsaros}}, \bibinfo {author} {\bibfnamefont {P.}~\bibnamefont {Spathis}},
  \bibinfo {author} {\bibfnamefont {M.}~\bibnamefont {Stoffel}}, \bibinfo
  {author} {\bibfnamefont {F.}~\bibnamefont {Fournel}}, \bibinfo {author}
  {\bibfnamefont {M.}~\bibnamefont {Mongillo}}, \bibinfo {author}
  {\bibfnamefont {V.}~\bibnamefont {Bouchiat}}, \bibinfo {author}
  {\bibfnamefont {F.}~\bibnamefont {Lefloch}}, \bibinfo {author} {\bibfnamefont
  {A.}~\bibnamefont {Rastelli}}, \bibinfo {author} {\bibfnamefont {O.~G.}\
  \bibnamefont {Schmidt}}, \ and\ \bibinfo {author} {\bibfnamefont {S.~D.}\
  \bibnamefont {Franceschi}},\ }\href {\doibase 10.1038/nnano.2010.84}
  {\bibfield  {journal} {\bibinfo  {journal} {Nature Nanotech.}\ }\textbf
  {\bibinfo {volume} {5}},\ \bibinfo {pages} {458} (\bibinfo {year}
  {2010})}\BibitemShut {NoStop}%
\bibitem [{\citenamefont {Herrmann}\ \emph {et~al.}(2010)\citenamefont
  {Herrmann}, \citenamefont {Portier}, \citenamefont {Roche}, \citenamefont
  {Levy~Yeyati}, \citenamefont {Kontos},\ and\ \citenamefont
  {Strunk}}]{herrmann_carbon_2010}%
  \BibitemOpen
  \bibfield  {author} {\bibinfo {author} {\bibfnamefont {L.~G.}\ \bibnamefont
  {Herrmann}}, \bibinfo {author} {\bibfnamefont {F.}~\bibnamefont {Portier}},
  \bibinfo {author} {\bibfnamefont {P.}~\bibnamefont {Roche}}, \bibinfo
  {author} {\bibfnamefont {A.}~\bibnamefont {Levy~Yeyati}}, \bibinfo {author}
  {\bibfnamefont {T.}~\bibnamefont {Kontos}}, \ and\ \bibinfo {author}
  {\bibfnamefont {C.}~\bibnamefont {Strunk}},\ }\href {\doibase
  10.1103/PhysRevLett.104.026801} {\bibfield  {journal} {\bibinfo  {journal}
  {Phys. Rev. Lett.}\ }\textbf {\bibinfo {volume} {104}},\ \bibinfo {pages}
  {026801} (\bibinfo {year} {2010})}\BibitemShut {NoStop}%
\bibitem [{\citenamefont {Lee}\ \emph {et~al.}(2012)\citenamefont {Lee},
  \citenamefont {Jiang}, \citenamefont {Aguado}, \citenamefont {Katsaros},
  \citenamefont {Lieber},\ and\ \citenamefont
  {De~Franceschi}}]{lee_zero-bias_2012}%
  \BibitemOpen
  \bibfield  {author} {\bibinfo {author} {\bibfnamefont {E.~J.~H.}\
  \bibnamefont {Lee}}, \bibinfo {author} {\bibfnamefont {X.}~\bibnamefont
  {Jiang}}, \bibinfo {author} {\bibfnamefont {R.}~\bibnamefont {Aguado}},
  \bibinfo {author} {\bibfnamefont {G.}~\bibnamefont {Katsaros}}, \bibinfo
  {author} {\bibfnamefont {C.~M.}\ \bibnamefont {Lieber}}, \ and\ \bibinfo
  {author} {\bibfnamefont {S.}~\bibnamefont {De~Franceschi}},\ }\href {\doibase
  10.1103/PhysRevLett.109.186802} {\bibfield  {journal} {\bibinfo  {journal}
  {Phys. Rev. Lett.}\ }\textbf {\bibinfo {volume} {109}},\ \bibinfo {pages}
  {186802} (\bibinfo {year} {2012})}\BibitemShut {NoStop}%
\end{thebibliography}

%

\end{document}